\documentclass[%
reprint,
 amsmath,amssymb,
prf,
floatfix,
]{revtex4-2}

\usepackage{comment}

\usepackage{stmaryrd}

\usepackage{subfig}

\usepackage{xcolor}

\newcommand{\seb}[1]{\textcolor{blue}{\textbf{#1}}}

\usepackage{graphicx}
\usepackage{dcolumn}
\usepackage{bm}
\usepackage{hyperref}
\usepackage{soul}

\allowdisplaybreaks
\begin{document}

\title{Aggregation and disaggregation processes in clusters of particles: simple numerical and theoretical insights of the competition in 2D geometries}

\author{Louis-Vincent Bouthier}
\email{Corresponding author\\
louis.bouthier@mines-paristech.fr}
 \author{Romain Castellani}%
\affiliation{%
  Groupe CFL, CEMEF, Mines Paris, PSL Research University, 1 Rue Claude Daunesse, 06904 Sophia Antipolis, France
}%

\author{S\'ebastien Manneville}
\affiliation{ENSL, CNRS, Laboratoire de physique, F-69342 Lyon, France}
\affiliation{Institut Universitaire de France (IUF)}
%

\author{Arnaud Poulesquen}
\affiliation{CEA, DES, ISEC, DE2D, SEAD, LCBC, Université de Montpellier, Marcoule, France
}%

\author{Rudy Valette}
\author{Elie Hachem}
\affiliation{ Groupe CFL, CEMEF, Mines Paris, PSL Research University, 1 Rue Claude Daunesse, 06904 Sophia Antipolis, France}

\date{\today}

\begin{abstract}

Aggregation and disaggregation of clusters of attractive particles under flow are studied from numerical and theoretical points of view. Two-dimensional molecular dynamics simulations of both Couette and Poiseuille flows highlight the growth of the average steady-state cluster size as a power law of the adhesion number, a dimensionless number that quantifies the ratio of attractive forces to shear stress. Such a power-law scaling results from the competition between aggregation and disaggregation processes, as already reported in the literature. Here, we rationalize this behavior through a model based on an energy function, which minimization yields the power-law exponent in terms of the cluster fractal dimension, in good agreement with the present simulations and with previous works.
\end{abstract}

\maketitle


\section{Introduction}


Colloidal suspensions composed of attractive particles, that aggregate into clusters when dispersed into a liquid medium, have raised great interest for a number of decades due to their wide range of applications, from paints and coatings to food products or construction materials \cite{Macosko1994,Mewis2012,Wagner2021,Hengl2014,Ioannidou2016,Awad2012,Knorr2004,Chandrapala2012}.
The large variety of attractive forces driving cluster aggregation, including van der Waals interactions, depletion or capillary forces, combined with stabilizing repulsive forces, such as electrostatic or steric interactions, lead to a complex phase diagram involving equilibrium liquid or phase-separated regimes as well as out-of-equilibrium gel or glassy phases \cite{Bonn2017,Mewis2012,Wagner2021,Vassileva2005}.  
Beyond their phase behaviour at rest, understanding how colloidal suspensions respond to deformation and flow is pivotal for their processing and their practical use. Thus, a huge amount of work has been devoted to the rheology of colloidal suspensions and their structure under external mechanical solicitations \cite{Mewis2012,Wagner2021,Macosko1994,Wessel1992,Shih1990}. In particular, one key question concerns the evolution of particle clusters under shear.

Colloidal clusters generically show a ramified structure  characterized by a fractal dimension $D$ linking the number of particles $n(l)$ in the cluster to its size $l$ through $n(l)\propto l^D$ \cite{Sorensen2001,Lin1990a,Lazzari2016}. Other structural parameters may be identified, such as the chemical dimension related to the number of bonds in the shortest path in a cluster \cite{Grassberger1985}, 
the connectivity or the coordination number \cite{Bantawa2021}. Since the seminal work of Weitz {\it et al.} \cite{Weitz1984,Weitz1985} on diffusion-limited aggregation, a large number of experimental studies have characterized the microscopic mechanisms underpinning cluster formation at rest, either from direct visualisation \cite{Hoekstra:2003,Nguyen2011}, or from scattering techniques  \cite{Richards2017}. In cases where colloidal aggregation leads to the formation of a space-spanning cluster network, i.e. to a colloidal gel, these measurements have provided key information on both the maximum cluster size $\ell$ and their fractal dimension $D$ \cite{Masschaele2009}. When shear is applied to a flocculated suspension, aggregates break up and form a suspension of isolated clusters. The steady-state cluster size distribution is a key observable and the dependence of $\ell$ and $D$ on the shear stress applied to the suspension has been probed by combining light, x-ray, and neutron scattering and rheometry \cite{Hoekstra:2005,Zaccone2009a,
Torres1991a,Hunter1980}. Typical results are that (i)~$\ell$ decreases as a power-law of the shear stress $\sigma$, $\ell\propto \sigma^{-m}$, with an exponent $m$ ranging from 0.2 to 0.5 \cite{Higashitani1998,Higashitani2001}, and (ii)~$D$ increases from 1.8--2.2 depending on the specific aggregation mechanism to 2.4--2.7 under shear, a phenomenon known as ``shear-induced cluster densification'' \cite{Hoekstra:2005,Massaro2020}.

Furthermore, thanks to the growth of computational capabilities over the past decade, large-scale numerical simulations have  been developed to model suspensions, both at rest and under shear, with up to several millions of colloidal particles \cite{Brady:1985,Morris:2009}, sometimes taking into account the solvent through hydrodynamic interactions \cite{Varga2015a,Varga2015b,Varga2018,Varga2019,Lorenzo2022}. The model suspensions rely on various idealized two-body interaction potentials \cite{Jungblut2019}, which in some cases, include bending stiffness \cite{Colombo2014a,Bouzid2018a,Bantawa2021}, or solid friction \cite{Kimbonguila2014,Eggersdorfer2010,Ruan2020}. 
One of the main advantages of numerical simulations is that they yield a complete picture of the cluster microstructure. Another modelling approach referred to as ``population balance model'' (PBM), relies on kinetic equations for ``classes'' of clusters of given sizes, which mimic the competition between aggregation and disaggregation processes through the use of various kernels \cite{Soos2006,Soos2007,Puisto2012,Lattuada2016}. Although some ingredients of the kernels can be theoretically predicted thanks to Smoluchowksi equations, in practice, PBM often require a significant amount of empirical tuning of the kernel parameters \cite{Lazzari2016,Banasiak2020a}.


Strong theoretical predictions have also been derived from analytical models that relate microscopic parameters, such as the interparticle bond stiffness, to macroscopic quantities, such as the elastic modulus $G'$ at rest, the yield strain $\gamma_y$ or the dynamic viscosity $\eta$ under shear \cite{Kantor1984a,Shih1990,Mellema2002}. In particular, one popular model by Wessel \& Ball \cite{Wessel1992} predicts that $\ell\propto\sigma^{-1/3}$ from a force balance under the assumption that the clusters behave hydrodynamically like compact spheres. While this model provides a correct estimate of the exponent $m$ for various experiments \cite{Torres1991a}, it does not account for the range of exponents reported in the literature \cite{Brakalov1987,Torres1991b,Hunter1980}, most probably due to the stringent assumption of hydrodynamically compact clusters. More refined models have considered soft, highly deformable clusters leading to $m=1/2$ rather than $m=1/3$ for rigid aggregates \cite{Snabre1996}. Yet, to the best of our knowledge, there is no general theoretical framework that may describe the behaviour of colloidal clusters under an external stress.

The goal of the present paper is to propose very general theoretical arguments to predict the steady-state size $\ell$ of colloidal aggregates submitted to an external solicitation. Starting from dimensional analysis, we show in Section~\ref{sec:framework} that the competition between cluster aggregation and disaggregation can be captured through the minimization of an energy function. In Section~\ref{sec:Numerical}, we then provide evidence for the existence of an energy minimum in simple numerical simulations. We proceed to detail an analytical model in Section~\ref{sec:Model}, which yields a power-law for $\ell$ as a function of the adhesion number, a dimensionless number that quantifies the ratio of attractive forces to shear stress. Finally, this model is compared to the coagulation-fragmentation approach and to previous experimental and numerical findings in Section~\ref{sec:Discussions}. Conclusions and open questions are drawn in Section~\ref{sec:Conclusion}.



\section{General framework}
\label{sec:framework}

Before describing the general approach based on a grand-canonical free energy, we start with simple dimensional arguments to justify that the competition between cluster aggregation led by attractive forces and disaggregation driven by shear forces depends on a single dimensionless group, namely the adhesion number $\mathrm{Ad}$, once the attraction range is fixed.
As discussed in Refs.~\cite{Kimbonguila2014,Marshall2014},
attractive forces may be estimated by $Ua/\delta^2$, with $U$ the depth of the interaction potential between two particles, $a$ the particle radius, and $\delta$ the range of interaction, which can be taken as the center-to-center distance between two particles at equilibrium or as the width of the potential well. Disaggregating forces, on the other hand, may be estimated by $\sigma a^2$, with $\sigma$ the external stress exerted on the clusters. One chooses here, as a distinction from the literature, to consider the stress rather than the shear rate because (i) whenever the shear rate is involved in similar definitions, it appears as multiplied  by the solvent viscosity, therefore as a shear stress, and (ii) one is convinced that stress drives the disaggregation, like in plasticity or fracture \cite{vonMises1913,Griffith1921,Irwin1957,Creton2016}, rather than a kinematic quantity. Following Ref.~\cite{Marshall2014}, the adhesion number is defined as the ratio of attractive forces to disaggregating forces:
\begin{equation}
\mathrm{Ad}=\frac{U}{\sigma a\delta^{2}}\,.\label{eq:AdhesionDefinition}
\end{equation}
The condition $\mathrm{Ad}\ll1$ implies that hydrodynamic forces dominate, while $\mathrm{Ad}\gg1$ indicates that attractive forces are predominant. It is important to note that the choice of this adhesion number contains some degree of arbitrariness. Indeed, other similar dimensionless groups may be built more generally by replacing $a\delta^2$ by $a^\alpha\delta^{3-\alpha}$, with $\alpha\in\left[0,3\right]$, in Eq.~(\ref{eq:AdhesionDefinition}). The precise choice depends on whether one considers energies ($\alpha=0$), forces ($\alpha=1$), or stiffnesses ($\alpha=2$), i.e., energies per unit surface, or energies per unit volume ($\alpha=3$).
Finally, according to the Vaschy-Buckingham theorem \cite{Buckingham1914,Buckingham1915a,Buckingham1915b}, any characteristic length that depends on $U$, $\sigma$, $a$, and $\delta$, such as the maximum cluster size $\ell$, may be expressed as $\ell/a=\mathcal{F}\left(\mathrm{Ad},\delta/a\right)$. Therefore, a combination of both $\mathrm{Ad}$ and $\delta/a$ is expected. Also note that the adhesion number simply corresponds to the inverse of the ``Mason number,'' $\mathrm{Mn}=\sigma a^2\delta/U$, a dimensionless group popular in the rheology community that quantifies the ratio of shearing forces to attractive forces \cite{Varga2019,Jamali2020,Nabizadeh2021}.

In order to describe the competition between aggregation and disaggregation, we consider a grand-canonical ensemble with a population of clusters of mass $k\in\mathbb{N}^*$ associated to a number of clusters $n_k$. Each cluster of mass $k$ has an associated energy $\mathcal{E}\left(k\right)$, which we seek to determine. The number of primary particles is not fixed and is related, for each cluster mass $k$, to a chemical potential $\alpha_k/\beta$ with $\beta=1/(k_B T)$, $k_B$ the Boltzmann constant, and $T$ the temperature. Note that the use of a canonical ensemble may seem more appropriate for a problem with a fixed number of particles. However, the calculation of the canonical partition function leads to the use of the complete Bell polynomials, where each variable is $\mathrm{e}^{-\beta\mathcal{E}\left(k\right)}$. Inverting the relation is not straightforward, so that computing the distribution of clusters of mass $k$ is very cumbersome within a canonical framework. Here, thanks to the grand-canonical formulation, $\mathcal{E}\left(k\right)$, as well as the energy $n_k\mathcal{E}\left(k\right)$ associated to all clusters of mass $k$, may be computed rather easily. The grand canonical partition function reads: 
\begin{align}
    \Xi&=\sum_{\left(n_{k}\right)_{k\in\mathbb{N}^*}\in\mathbb{N}^{\mathbb{N}^*}}\exp\left(-\sum_{k\in\mathbb{N}^*}\left(\beta n_{k}\mathcal{E}\left(k\right)-n_{k}k\alpha_{k}\right)\right)\label{eq:series}\\
    &=\prod_{k\in\mathbb{N}^*}\left(1-\mathrm{e}^{-\beta\mathcal{E}\left(k\right)+k\alpha_{k}}\right)^{-1}.
\end{align}
The convergence of the series in Eq.~(\ref{eq:series}) is guaranteed if $\alpha_kk<\beta\mathcal{E}\left(k\right)$ for all $k\in\mathbb{N}^*$. Moreover, it follows from Eq.~(\ref{eq:series}) that the distribution of each level population $n_k$ is a geometric distribution with a parameter $\mathrm{e}^{-\beta\mathcal{E}\left(k\right)+k\alpha_k}$. Since the average number of particles $\left\langle N\right\rangle$ is related to the average population $\left(\left\langle n_k\right\rangle\right)_{k\in\mathbb{N}^*}$ of clusters of mass $k$ through $\left\langle N\right\rangle=\sum_{k\in\mathbb{N}^*}k\left\langle n_k\right\rangle$, one may compute the average number of clusters of mass $k$ and the standard deviation through: 
\begin{align}
    \left\langle n_k\right\rangle&=-\frac{1}{k}\frac{\partial \ln \Xi}{\partial\alpha_k}=\left(\mathrm{e}^{\beta\mathcal{E}\left(k\right)-\alpha_{k}k}-1\right)^{-1},\\
    \Delta n_k&=\frac{1}{k}\sqrt{\frac{\partial^2\ln \Xi}{\partial\alpha_k^2}}=\sqrt{\left\langle n_k\right\rangle\left(1+\left\langle n_k\right\rangle\right)}\\
    &=\frac{1}{2}\,\mathrm{csch}\left(\frac{\beta\mathcal{E}\left(k\right)-\alpha_{k}k}{2}\right),
\end{align}
which boils down to the Bose-Einstein statistics.

In practice, in an experiment or a simulation, a initial sample of $\left(n_k\right)_{k\in\mathbb{N}^*}$ is chosen, which evolves in time according to external conditions. It is clear from the above expressions that the $\left(n_k\right)_{k\in\mathbb{N}^*}$ should end up being centered around the averages $\left(\left\langle n_k\right\rangle\right)_{k\in\mathbb{N}^*}$ with a dispersion $\left(\Delta n_k\right)_{k\in\mathbb{N}^*}$. Therefore, a higher number of clusters of mass $k$ also brings a higher dispersion around this value. 
Since numerical simulations yield a probability distribution function $f\left(k\right)$ for the number of cluster of mass $k$ , which is assumed to be equal to the average distribution $\left\langle n_k\right\rangle$, i.e., $f\left(k\right)=\left\langle n_k\right\rangle$, one can find the average potential energy of the clusters and the standard deviation respectively through:
\begin{align}
    \beta\mathcal{E}\left(k\right)-\alpha_{k}k&=\ln\left(1+\frac{1}{f\left(k\right)}\right)\label{eq:Energy}\\
    \Delta\left(\beta\mathcal{E}\left(k\right)-\alpha_{k}k\right)&=\left(f\left(k\right)\left(1+f\left(k\right)\right)\right)^{-\frac{1}{2}}\\
    &=2\,\sinh\left(\frac{\beta\mathcal{E}\left(k\right)-\alpha_{k}k}{2}\right)\\
    &>\beta\mathcal{E}\left(k\right)-\alpha_kk\,.
\end{align}

\section{Numerical simulation}

\label{sec:Numerical}

\subsection{Numerical scheme and analysis}

To get some insight of the competition between aggregation and disaggregation processes, we turn to simple molecular dynamics simulations of two-dimensional Couette and Poiseuille flows based on the LAMMPS library \cite{Thompson2022}. We use reduced units, where the unit size is given by the particle radius. The simulation box is of size $\mathcal{L}\times h$ with periodic boundary conditions along the $x$ direction and solid boundaries along the $y$ direction located at $y=0$ and $y=h$. The length of the box $\mathcal{L}$ is fixed to $\mathcal{L}=620$, and the width $h$ is either 103 or 206, much larger than the particle size. 
Each boundary is constituted of one layer of particles of unit size which positions are fixed and that interact with bulk particles through a repulsive Yukawa potential $w_s(r)= 100\mathrm{e}^{-r}/r$, with $r$ the distance between two particles and a cut-off distance $r=5$ to save computation time. The interaction potential between two particles in the bulk is a classical 12-6 Lennard-Jones potential $w(r)=4\left(r^{-12}-r^{-6}\right)$ with a cut-off distance $r=40$, again to save computation time.

First, the system is initialised with a particle surface fraction of $\phi=0.12$. Using a canonical formulation, the temperature is fixed at $T=0.01$. The time step is taken as $\Delta t=0.01$, which is small enough to account realistically for temporal variations while keeping the computation time reasonably low. Particles are distributed over a square lattice, and their initial velocities are chosen according to a Maxwell distribution for the given temperature. The system is then let to evolve for a duration $5,000$ time units in order to create the initial cluster distribution. More precisely, the equation of motion for each particle $i\in\left\llbracket1,N\right\rrbracket$, with $N$ the total number of particle, is 
\begin{align}
    \frac{\mathrm{d}^2\boldsymbol{r}_i}{\mathrm{d}t^2}&=-\frac{\partial \mathcal{W}}{\partial \boldsymbol{r}_i}\left(\left(\boldsymbol{r}_j\right)_{j\in\left\llbracket1,N\right\rrbracket}\right)+\boldsymbol{B}_i\\
    \mathcal{W}\left(\left(\boldsymbol{r}_j\right)_{j\in\left\llbracket1,N\right\rrbracket}\right)&=\sum_{k=1}^N\sum_{j=k+1}^Nw\left(\left|\boldsymbol{r}_k-\boldsymbol{r}_j\right|\right)\\
    w\left(r\right)&=4\left(r^{-12}-r^{-6}\right)\,,
\end{align}
with $\boldsymbol{r}_i$ the position of particle $i$ and $\boldsymbol{B}_i$ a Brownian white noise for particle $i$.

\begin{figure*}
    \centering
    \includegraphics[width=1\columnwidth]{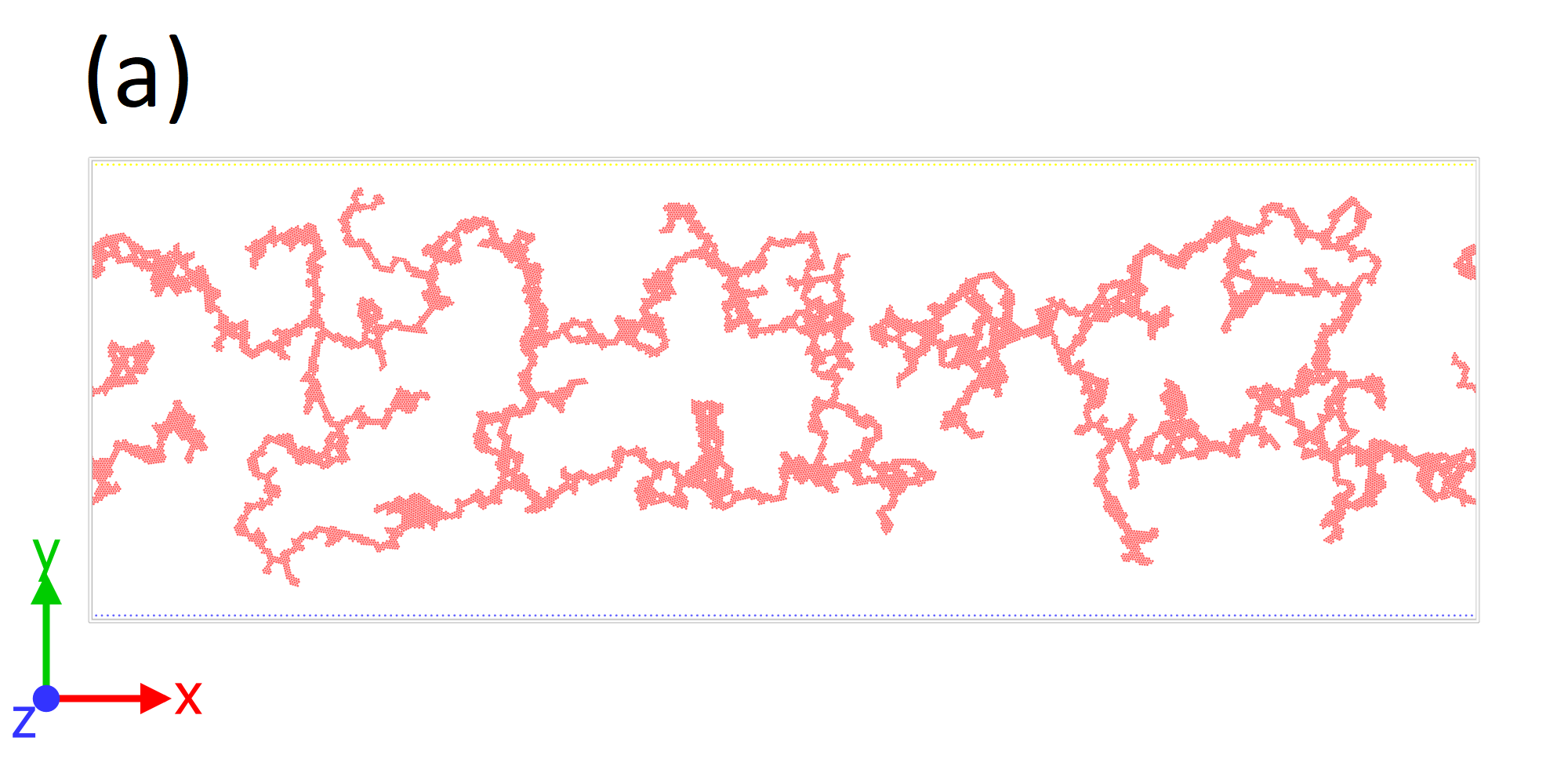}\includegraphics[width=1\columnwidth]{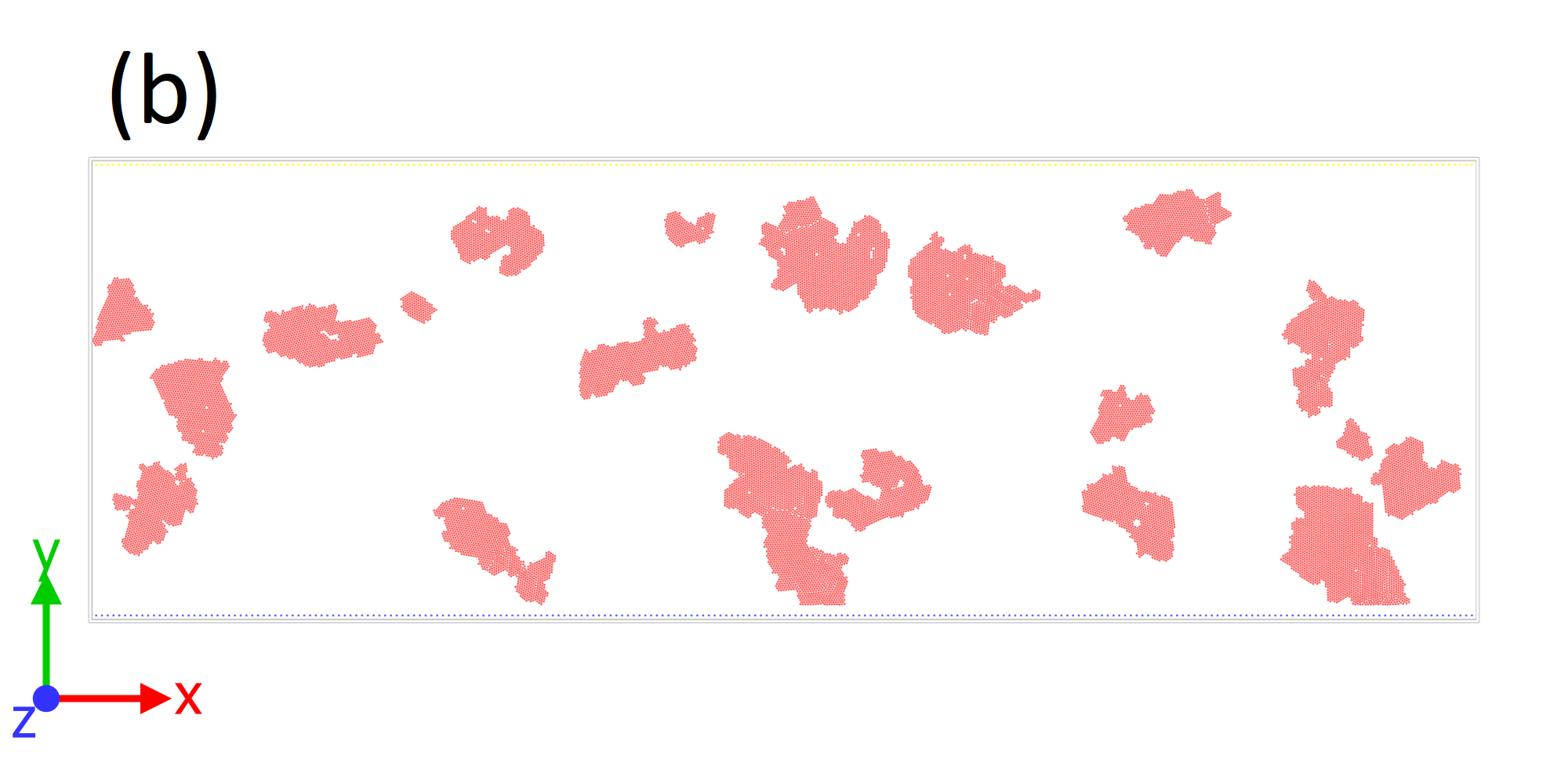}
    \includegraphics[width=2\columnwidth]{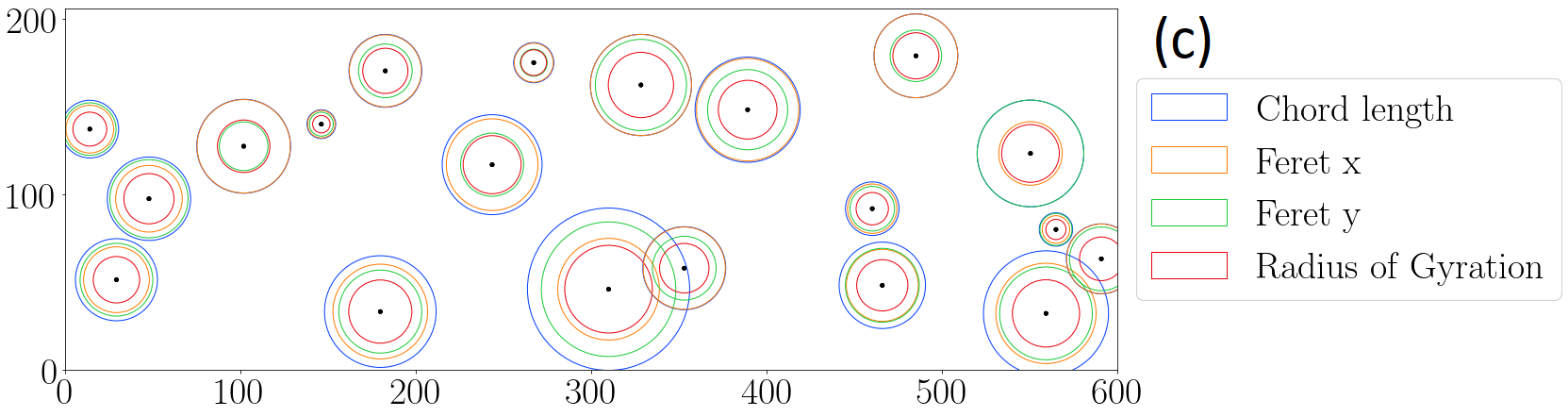}
    \caption{Simulated particle distributions for $V=10$, $h=206$, and $C=1$ after the preparation step prior to shearing (a) and at after shearing in the Couette geometry (b). Rendering using \emph{Ovito} \cite{Stukowski2010}. (c) Cluster detection and various estimates of the cluster size as defined in the text for the top right particle distribution.}
    \label{fig:VonMises}
\end{figure*}

In a second step, a drag force $\boldsymbol{F}_i=C\left(\boldsymbol{u}-\frac{\mathrm{d}\boldsymbol{r}_i}{\mathrm{d}t}\right)$ is applied on each particle $i\in\left\llbracket1,N \right\rrbracket $, with $C=1$ the drag coefficient \footnote{Additional simulations were carried out with various values of $C$ and the main results remained similar up to a rescaling of the time scale.}
, $\boldsymbol{u}=Vy/h\boldsymbol{e}_x$ for the 2D Couette flow or $\boldsymbol{u}=4V\left(1-y/h\right)y/h\boldsymbol{e}_x$ for the 2D Poiseuille flow, where $V$ is the maximum flow velocity. Here, a micro-canonical formulation is used and the time step is set to $\Delta t=0.001$. The simulation is run for a duration of $1,000$ time units. The equation of motion for each particle $i$ thus reads: 
\begin{align}
    \frac{\mathrm{d}^2\boldsymbol{r}_i}{\mathrm{d}t^2}&=-\frac{\partial \mathcal{W}}{\partial \boldsymbol{r}_i}\left(\left(\boldsymbol{r}_j\right)_{j\in\left\llbracket1,N\right\rrbracket}\right)+F\left(\boldsymbol{r}_i\right)\boldsymbol{e}_x\\
    F\left(\boldsymbol{r}_i\right)&=C\left(-\frac{\mathrm{d}\boldsymbol{r}_{i}}{\mathrm{d}t}\cdot\boldsymbol{e}_{x}+\right.\\
    &\left.V\frac{\boldsymbol{r}_{i}\cdot\boldsymbol{e}_{y}}{h}\begin{cases}
1 & \text{2D Couette flow}\\
4\left(1-\frac{\boldsymbol{r}_{i}\cdot\boldsymbol{e}_{y}}{h}\right) & \text{2D Poiseuille flow}
\end{cases}\right)\,. 
\end{align}

The parameters investigated in the present work are $\left(V,h\right)\in\left\{0.1,0.3,1,3,10,30,100\right\}\times\left\{103,206\right\}$. The numerical scheme used here is a Velocity Verlet algorithm \cite{Swope1982}. The computations output are (i)~the position of the particles at each time step, and (ii)~the clusters to which the particles belong based on a connected-component algorithm \cite{Pearce2005} with a distance threshold of 1.4,
consistently with the literature \cite{Colombo2013,Colombo2014a,Colombo2014b}. Moreover, we checked that for thresholds ranging from 1.1 to 2.0, the distribution of neighbours does not change, so that the results are not sensitive to the specific choice of threshold. The size of cluster number $I$ is quantified according to the following estimates:
\begin{itemize}
    \item the radius of gyration $R_g^I$ given by 
    \begin{align}
        R_g^I&=\sqrt{\frac{1}{\left|P\left(I\right)\right|}\sum_{k\in P\left(I\right)}\left\lVert\boldsymbol{r}_k-\overline{\boldsymbol{r}}\right\rVert^2},\\
        \overline{\boldsymbol{r}}&=\frac{1}{\left|P\left(I\right)\right|}\sum_{k\in P\left(I\right)}\boldsymbol{r}_k,
    \end{align}
    with $\boldsymbol{r}_k$ the position of the particles in the cluster, $P\left(I\right)$ the set of particles in cluster $I$ and $\left|P\left(I\right)\right|$ the number of particles in cluster $I$.  
    \item the Feret radii in the $x$ and $y$ directions given by $\left(\max_{k\in P\left(I\right)}\boldsymbol{r}_k\cdot\boldsymbol{e}\right)/2-\left(\min_{k\in P\left(I\right)}\boldsymbol{r}_k\cdot\boldsymbol{e}\right)/2$ with $\boldsymbol{e}$ the unit vector in the $x$ and $y$ directions respectively.
    \item the half maximum chord length given by $\max_{\left(k,l\right)\in P\left(I\right)^2}\left\lVert\boldsymbol{r}_k-\boldsymbol{r}_l\right\rVert/2$.
\end{itemize}
In order to infer statistical estimations,  each size distribution is further weighted by the number of particles in each cluster. Such weighting is needed because the number of clusters is not constant.  Therefore, because the total number of particles is constant, weighting by the mass of each cluster allows one to recover the number of particles when integrating over the whole distribution. Finally, thanks to the reduced units, $U/a\delta^2=1$ and the shear stress is $\sigma=CV/h=V/h$ here due to $C=1$ so that the adhesion number simply reads $\mathrm{Ad}=h/V$.

\subsection{Simulation results}
\label{sec:SimRes}

Figure~\ref{fig:VonMises} shows typical particle distributions computed after the preparation step prior to shearing (a), and after application of shear in the Couette geometry (b). It appears clearly that the system starts from a space-spanning ramified structure and evolves toward dense, isolated clusters under shear. 
Moreover, as shown in Fig.~\ref{fig:VonMises}(c), the different estimates for the cluster size yield consistent values. In the following, for the sake of simplicity, we shall focus only on the weighted average of the radius of gyration to estimate the cluster size $\ell$.  

\begin{figure*}
    \centering
    \includegraphics[width=2\columnwidth]{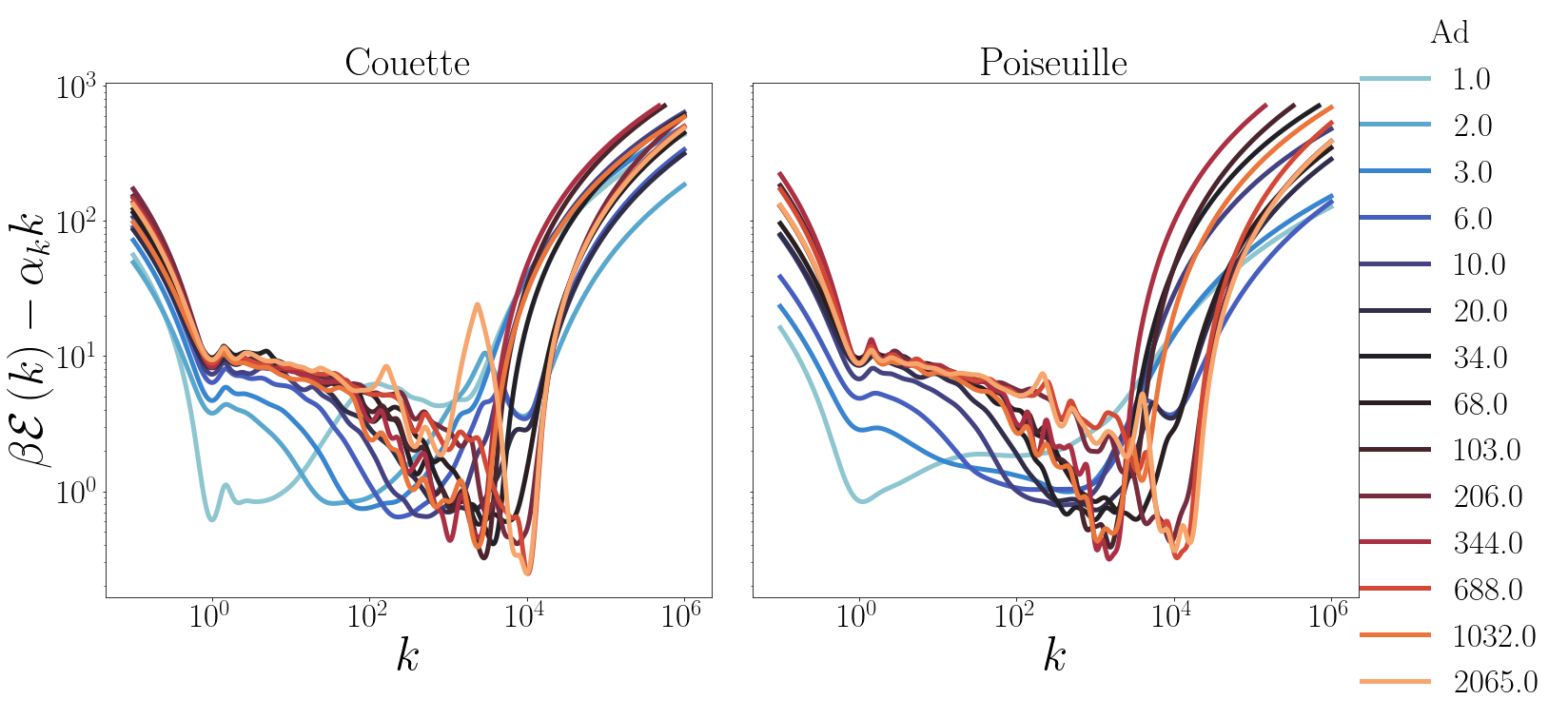}
    \caption{Energy functional $\beta\mathcal{E}\left(k\right)-\alpha_kk$ computed from from Eq.~(\ref{eq:Energy}) as a function of the number $k$ of particles in a cluster for (left) Couette flow and (right) Poiseuille flow. Colors correspond to the adhesion number $\mathrm{Ad}$ as indicated in the legend. Results obtained from simulations performed with $h=103$ or $h=206$.}
    \label{fig:Energy}
\end{figure*}

Following the general framework introduced in Sect.~\ref{sec:framework}, we compute the distribution of the cluster mass and the dimensionless energy $\beta\mathcal{E}\left(k\right)-\alpha_kk$ based on Eq. (\ref{eq:Energy}). Figure~\ref{fig:Energy} shows this energy functional plotted against the mass $k$ of particles within a cluster for adhesion numbers ranging from 1 to about $2,000$. First, looking at the small- and large-mass limits, it appears that $\lim_{k\to0^+}\beta\mathcal{E}\left(k\right)-\alpha_kk=\lim_{k\to+\infty}\beta\mathcal{E}\left(k\right)-\alpha_kk=+\infty$, which indicates that extreme masses are not accessible to the system. 
Second, in all cases, there exists a global minimum of the energy functional that shifts towards larger values of $k$ as the adhesion number $\mathrm{Ad}$ is increased. As expected intuitively, this suggests that the average cluster mass increases with $\mathrm{Ad}$. 
Moreover, the steeper slope $\partial_k\left(\beta\mathcal{E}\left(k\right)-\alpha_kk\right)$ of the energy functional on the right side of the global minimum than on the left side indicates that the system reaches the energy minimum more rapidly when starting from large masses than from small masses. This confirms the intuition that disaggregation processes are much faster than aggregation processes. 
Third, while the value of the global minimum energy does not show any clear trend with $\mathrm{Ad}$,
there may exist several local minima in $\beta\mathcal{E}\left(k\right)-\alpha_kk$. This means that several metastable states may occur. These states may disappear when increasing the simulation duration or the system size. Still, this shows that a rather polydisperse population of clusters may be found, at least transiently. This may also be related to the dispersion in the energy minimum illustrated previously. 

\begin{figure*}
    \centering
    \includegraphics[width=1.3\columnwidth]{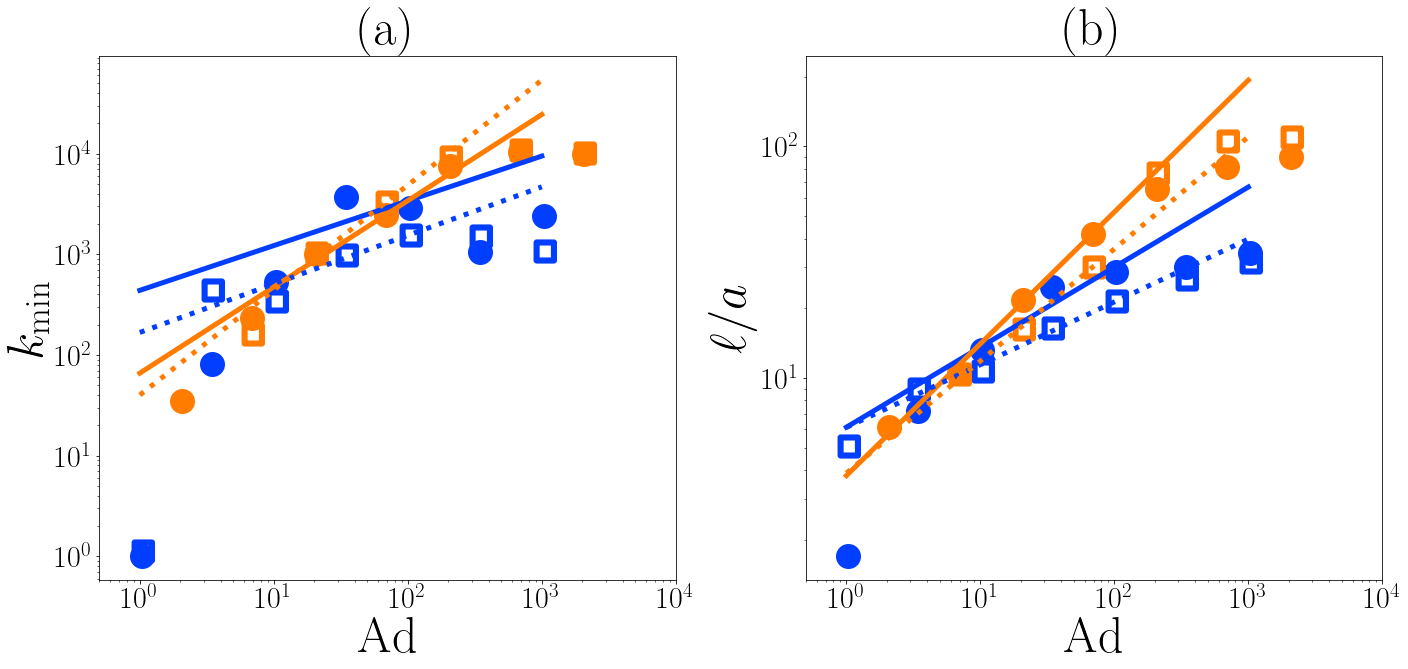}
    \includegraphics[width=0.57\columnwidth]{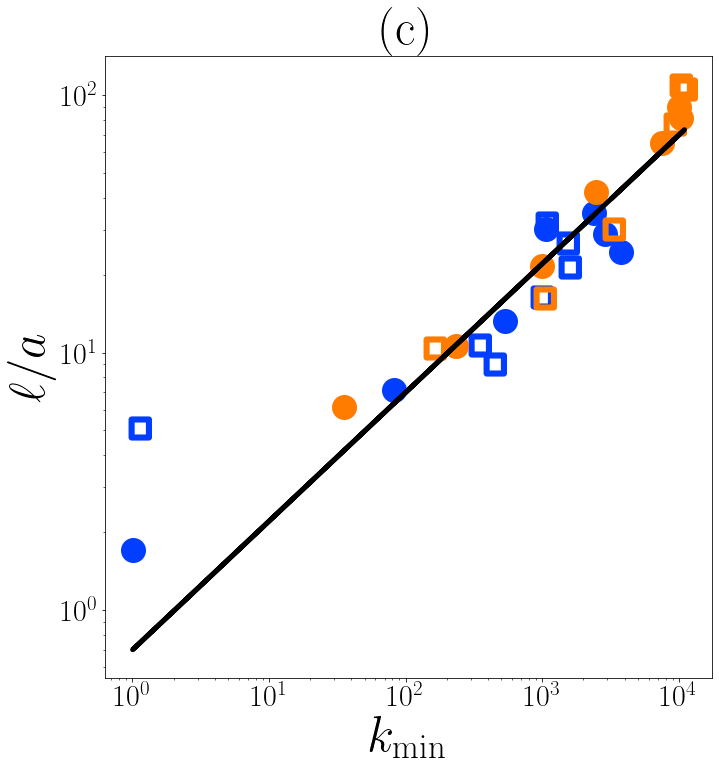}
    \caption{(a) Mass $k_{\min}$ of the clusters corresponding to the energy minimum and (b)~cluster size $\ell$ defined as the weighted average of the radius of gyration of the clusters, as a function of the adhesion number $\mathrm{Ad}$ for Couette flow ($\bullet$ ) and Poiseuille flow ($\square$) and for two different width $h=103$ (blue) and $h=206$ (orange). Solid and dotted lines correspond to the best power-law fits respectively for the Couette and the Poiseuille flow, and computed over $\mathrm{Ad}\in\left[2,200\right]$. (c)~Cluster size $\ell$ as a function of the equilibrium mass $k_{\min}$. The black solid line corresponds to $\ell\propto k_{\min}^{1/2}$. Same symbols and colors as in (a) and (b).
   }
    \label{fig:Size_kmin_Ad2}
\end{figure*}

\begin{table}[]
    \centering
    \begin{tabular}{ccccc}
         \hline
         Geometry & $h$ & Variable & Exponent & Prefactor \\
         \hline
         Couette & 103 & $\ell/a$ & $0.346\pm0.009$ & $6.1\pm0.2$ \\
         Couette & 103 & $k_{\min}$ & $0.4\pm0.4$ & $400\pm700$\\
         Couette & 206 & $\ell/a$ & $0.57\pm0.03$ & $3.5\pm0.4$ \\
         Couette & 206 & $k_{\min}$ & $0.9\pm0.1$ & $60\pm30$\\
         Poiseuille & 103 & $\ell/a$ & $0.271\pm0.007$ & $6.1\pm0.2$ \\
         Poiseuille & 103 & $k_{\min}$ & $0.5\pm0.1$ & $170\pm70$\\
         Poiseuille & 206 & $\ell/a$ & $0.48\pm0.03$ & $3.9\pm0.5$ \\
         Poiseuille & 206 & $k_{\min}$ & $1.0\pm0.1$ & $40\pm20$\\
         \hline
    \end{tabular}
    \caption{Exponents and prefactors of the power-law fits in Fig. \ref{fig:Size_kmin_Ad2} for the different geometries, widths, and sizes $\ell$ and $k_{\min}$.}
    \label{tab:ExponentsFactor2}
\end{table}

Figure~\ref{fig:Size_kmin_Ad2}(a) and Fig.~\ref{fig:Size_kmin_Ad2}(b) respectively show the ``equilibrium mass'' $k_{\min}$ of the clusters corresponding to the energy minimum, i.e., $k_{\min}=\arg\min_{k\in\mathbb{R}_+}\left(\beta\mathcal{E}\left(k\right)-\alpha_kk\right)$, and the cluster size $\ell/a$ taken as the weighted average of their radius of gyration plotted against the adhesion number $\mathrm{Ad}$. For both observables, a power-law regime is identified over almost two decades in adhesion numbers. The exponents inferred from power-law fits for $\mathrm{Ad}\in\left[2,200\right]$ are reported in Table~\ref{tab:ExponentsFactor2}. For $\mathrm{Ad}\gtrsim 100$, a saturation is observed in $k_{\min}$ together with large variations, for both the Couette and Poiseuille flows. This behavior is most probably linked to finite-size effects as the average cluster size becomes comparable or larger than the system width $h$. Consistently with Fig.~\ref{fig:Size_kmin_Ad2}(a), the cluster size $\ell$ increases as a power-law of $\mathrm{Ad}$ [Fig.~\ref{fig:Size_kmin_Ad2}(b)]. There, although the data for $\ell$ do not show such a strong saturation as for $k_{\min}$, significant deviations from power-law behavior are still observed for $\mathrm{Ad}\gtrsim 100$.

Moreover, the exponents for the dependence of $\ell$ with $\mathrm{Ad}$ seem to depend significantly on the geometry, with values 0.35 and 0.57 for the Couette flow and 0.27 and 0.48 for the Poiseuille flow, respectively for $h=103$ and $h=206$ (see Table~\ref{tab:ExponentsFactor2}). Similarly, the corresponding exponents for $k_{\min}$, namely 0.4 and 0.9 for Couette flow, and 0.5 and 1.0 for Poiseuille flow, also differ for the two values of $h$. At this stage, the reason for such an influence of $h$ in the simple two-dimensional simulations remains unclear, and future work should focus on more realistic interaction potentials and three-dimensional geometries. 

Still, it is interesting to note that $\ell/a$ and $k_{\min}$ are not expected to have the same power-law behavior with the adhesion number because of their intrinsic relationship. More specifically, considering the framework of fractal clusters, the number of particles in a cluster $k$ is related to the geometrical size $l$ through $k=\left(l/a\right)^D$ with $D$ the fractal dimension. In the two-dimensional case, $D$ should fall into the range $\left[1,2\right]$. Figure~\ref{fig:Size_kmin_Ad2}(c) shows that $\ell/a\propto k_{\min}^{1/2}$, which is compatible with a fractal dimension $D\simeq 2$, i.e., with almost dense aggregates in two dimensions geometries. The fact that very compact aggregates are obtained in the simulations is confirmed visually by looking at Fig.~\ref{fig:VonMises}(b). Another confirmation comes from the ratio of the exponent for $k_{\min}$ and that for $\ell/a$ in Table \ref{tab:ExponentsFactor2}, which is also found to be close to 2 (except for the Couette flow with $h=103$ but the scatter of the $k_{\min}$ data in this latter case makes it difficult to conclude). 

\section{Proposition of model}

\label{sec:Model}

\subsection{Description of the disaggregation and reaggregation processes}

\begin{figure}
    \centering
    \includegraphics[width=\columnwidth]{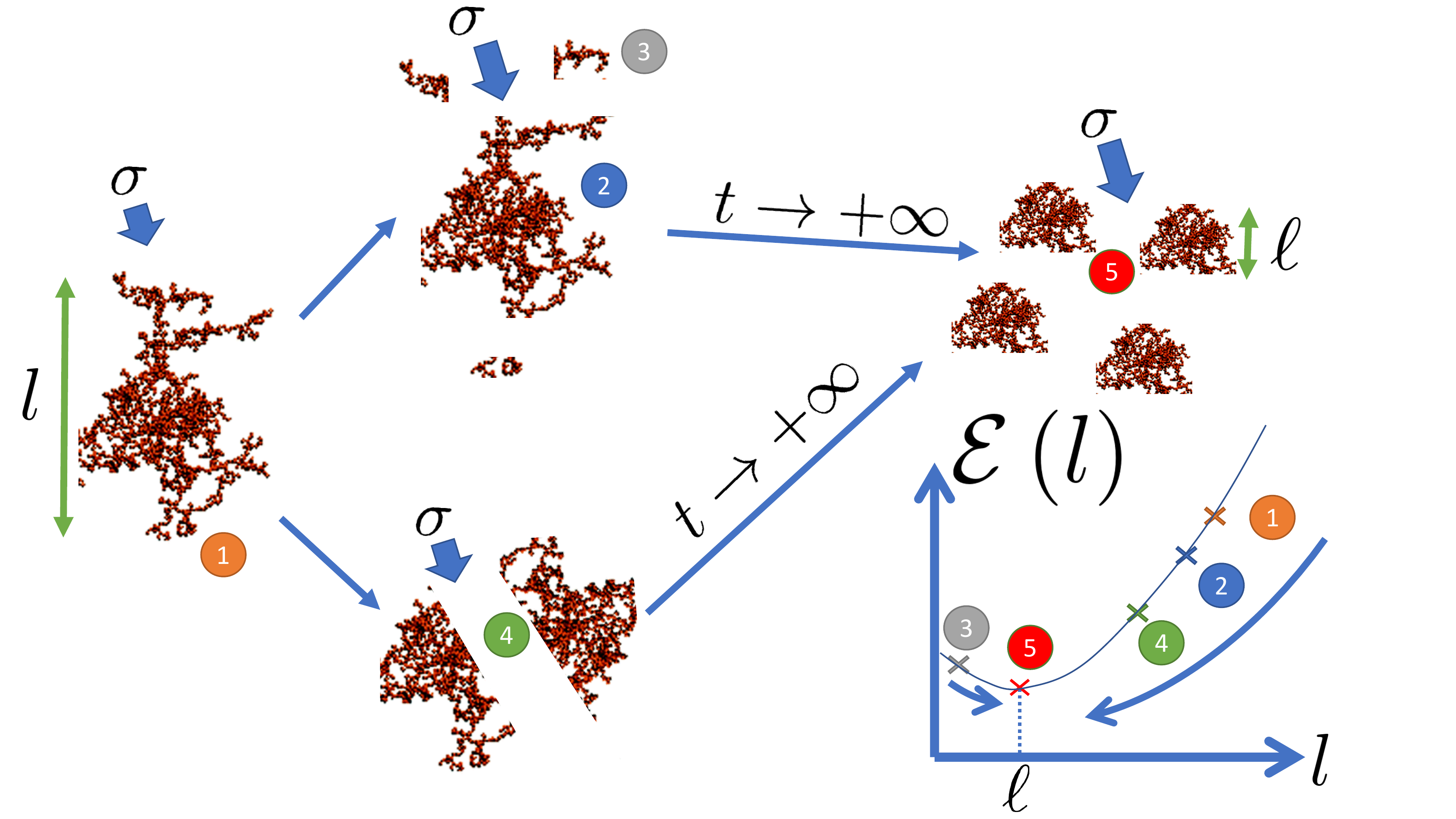}
    \caption{Sketch of the disaggregation and reaggregation processes under a uniform external stress $\sigma$. The numbers represent different sizes of aggregates with different levels of energy. The aggregates are disaggregating and reaggregating according to the stress solicitation. Transition from state 1 to states 2 and 4 corresponds to a ``fragile'' rupture, while transition from state 1 to state  3 is an ``erosion'' process.}
    \label{fig:SketchAggregates}
\end{figure}

The precise internal structure of the aggregates is really complex, and depends upon many different parameters including the volume fraction, the nature of the interparticle forces and of the solvent, temperature, and chemical environment. Hence, we shall consider the aggregates as a continuum without further internal details. 
We consider a suspension of particles of size
$a$ that interact through an attractive potential so that they gather into aggregates as sketched in Fig.~\ref{fig:SketchAggregates}. Following a statistical approach as in Refs.~\cite{Banasiak2020a,Banasiak2020b,Golse2005,Alexeev2004,Stadnichuk2015}, we assume that there exists a distribution $f\left(t,l,a,U,\delta,\sigma\right)$ giving the population of aggregates of size $l$ at time $t$ made of particles of size $a$ interacting through a pair potential $U$ over a distance $\delta$ under a stress solicitation $\sigma$. Such a distribution may be computed through a coagulation-fragmentation equation as in Refs.~\cite{Banasiak2020a,Banasiak2020b,Stadnichuk2015}. Yet, this approach requires to identify coagulation kernels and fragmentation coefficients, which involves much effort for theoretical and/or numerical validation. Without ignoring the power of such tools, we aim at a simpler approach through the present statistical approach. 

We start by imposing the conservation of the total number of particles $N$, which reads:
\begin{equation}
    N=\int_0^{+\infty}n\left(\frac{l}{a}\right)f\left(t,l,a,U,\delta,\sigma\right)\,\mathrm{d}l=\mathrm{constant}\,,
\end{equation}
where $n\left(l/a\right)$ denotes the number of particles per cluster of relative size $l/a$. 
The average size of the aggregates is thus given by:
\begin{equation}
    \overline{l}\left(t,a,U,\delta,\sigma\right)=\frac{a}{N}\int_0^{+\infty}n\left(\frac{l}{a}\right)\frac{l}{a}f\left(t,l,a,U,\delta,\sigma\right)\,\mathrm{d}l\,.
    \label{eq:lbar}
\end{equation}
From a statistical point of view, and assuming that the aggregates are submitted to a uniform external stress $\sigma$ far from any boundary,
aggregates should evolve from one state 
to another as sketched in Fig.~\ref{fig:SketchAggregates}. If the stress is sufficiently large to break some initial aggregate (state 1), a disaggregation occurs either due to
``fragile'' rupture into two pieces of similar sizes (bottom part of Fig.~\ref{fig:SketchAggregates}, state 4) or due to ``erosion,'' where small pieces detach from the initial aggregate
(top part of Fig.~\ref{fig:SketchAggregates}, states 2 and 3). 
If the larger pieces can still be broken down (states 2 and 4), the process continues. However, if the pieces become too small (state 3), the interparticle attraction dominates and reaggregation occurs. 
In the process, the aggregates progressively decrease 
their global energy, until a minimum is reached at long times
$t\to+\infty$. Therefore, the steady state (state 5) eventually corresponds to the optimum of all
possible sizes, which results from a dynamical equilibrium between disaggregation and reaggregation processes. More formally, when a steady state is reached, the average aggregate size $\ell$ is given by \begin{equation}
    \ell\left(a,U,\delta,\sigma\right)=\lim_{t\to+\infty}\overline{l}\left(t,a,U,\delta,\sigma\right). \label{eq:limitAverage}
\end{equation}
In practice, since $t$ remains finite, we note that the longer the final time, the smaller the spread of the distribution around the steady-state size. An additional comment is that, keeping the previous set of variables for the distribution $f$, an equivalent form using the Vaschy-Buckingham theorem \cite{Buckingham1914,Buckingham1915a,Buckingham1915b} is found by replacing $f\left(t,l,a,U,\delta,\sigma\right)\mathrm{d}l$ with $\tilde{f}\left(l/a,\mathrm{Ad},\delta/a\right)\mathrm{d}\left(l/a\right)$, where $t$ can be discarded based on unit independence. Time becomes relevant, however, when the viscosity $\eta$ of the suspending liquid or any other time-related quantity is considered.

\subsection{Analytical formulation of the model}
\label{subsec:Analytical}

Let us consider a suspension of particles of diameter
$a$ gathered in aggregates. The bond between
each particle involves an energy $U$ and an interparticle distance
of separation $\delta$. The system is submitted to a uniform stress
$\sigma$. One aims at estimating the steady-state size of stable aggregates $\ell$ as a function of $a$, $U$, $\delta$, and $\sigma$. Following \textcite{Eggersdorfer2010}, for a dense aggregate of size $l$, the applied mechanical energy per unit area is $\sigma l$. However, aggregates are not completely dense and one should more generally account for their fractal nature.
As already introduced above in Sect.~\ref{sec:SimRes}, the number of 
particles in an aggregate of size $l$ is proportional to $\left(l/a\right)^D$ with $D$ the fractal dimension of the clusters. 
Hence, on the one hand, the mechanical energy per unit of effective surface of the aggregate 
$M\left(l\right)$ is proportional to $\sigma a\left(l/a\right)^{D+1}$. On the other hand, if one
isolates an aggregate of size $l$, the energy per unit surface
liberated due to broken bonds $E$ is $E\left(l\right)=U\delta^{-2}$, which is independent of the size $l$ \cite{Marshall2014}.
Indeed, considering an intermolecular potential $w\left(d\right)$, with $d$ the intermolecular distance, one can compute the interparticle potential per unit area through:
\begin{equation}
    \label{eq:interaction}
    W\left(d\right)=2\pi\varrho^2\intop_d^{+\infty}\intop_y^{+\infty}\intop_0^{+\infty}rw\left(\sqrt{r^2+z^2}\right)\,\mathrm{d}r\mathrm{d}z\mathrm{d}y\,,
\end{equation}
where $\varrho$ is the molecular density inside a particle. Defining $\delta$ as the distance that satisfies $W'\left(\delta\right)=0$, or equivalently $\delta=\arg\min_{d\in\mathbb{R}_+^*} W\left(d\right), $\footnote{Taking some usual examples leads to $\delta\approx\arg\min_{d\in\mathbb{R}_+^*}w\left(d\right)$}, and the energy $U=\delta^2W\left(\mathfrak{\delta}\right)$, $E\left(l\right)=U\delta^{-2}$ indeed corresponds to the energy per unit surface liberated due to broken bonds. This energy is independent of the size of the aggregate because one may assume that, on the boundary of the aggregate, the number of particles per unit surface does not depend on $l$ and is only related to the structure, which is assumed to be fixed in steady state. Finally, the steady-state size $\ell$ corresponds to the size for which the mechanical energy balances that due to broken bonds, i.e., $M\left(\ell\right)=E\left(\ell\right)$, which leads to:
\begin{equation}
\frac{\ell}{a}=\left(\frac{U}{\sigma a\delta^{2}}\right)^{\frac{1}{1+D}}=\mathrm{Ad}^{\frac{1}{1+D}}.\label{eq:theory}
\end{equation}
The same result can be obtained by minimizing the total energy $\mathcal{E}\left(l\right)=l^2\left(M\left(l\right)-E\left(l\right)\right)$. Equation~(\ref{eq:theory}) also agrees with Eq. (\ref{eq:limitAverage}) through the Vaschy-Buckingham theorem \cite{Buckingham1914,Buckingham1915a,Buckingham1915b}. Indeed, Eq.~(\ref{eq:limitAverage}) may be rewritten as: 
\begin{equation}
    \frac{\ell}{a}=G\left(\mathrm{Ad},\frac{\delta}{a}\right).
\end{equation}
Adding that $U/\delta^2$ provides all the information about the interaction potential, one has $\partial_{\delta/a}G=0$ and thus 
\begin{equation}
    \frac{\ell}{a}=G\left(\mathrm{Ad}\right)\,,
\end{equation}
consistently with Eq.~(\ref{eq:theory}). 

\section{Discussion}

\label{sec:Discussions}

In this section, we discuss the theoretical approach and numerical results in light of the literature. We start by comparing the model with the classical coagulation-fragmentation approach. Then, the power-law scaling predicted for the cluster size, $\ell\propto \mathrm{Ad}^{1/(1+D)}$, is confronted to the present simulations and to previous experimental results. 

\subsection{Comparison with a coagulation-fragmentation model}

The model proposed above in Sect.~\ref{sec:Model} may be compared to the coagulation-fragmentation approach introduced in the literature more than thirty years ago \cite{Sorensen1987} and subsequently enriched over the years, e.g., through the population balance models \cite{Banasiak2020a,Banasiak2020b,Lattuada2016,Puisto2012}. In such an approach, the probability density function $f$ of clusters of size $x$ at time $t$ obeys the following dynamical equation:

\begin{multline}
    \frac{\partial f}{\partial t}\left(t,x,\bullet\right)=\\
    \frac{1}{2}\int_0^xK\left(y,x-y,\bullet\right)f\left(t,y,\bullet\right)f\left(t,x-y,\bullet\right)\,\mathrm{d}y\\
    -\frac{1}{2}\int_0^xF\left(y,x-y,\bullet\right)f\left(t,x,\bullet\right)\,\mathrm{d}y\\
    -\int_0^{+\infty}K\left(x,y,\bullet\right)f\left(t,x,\bullet\right)f\left(t,y,\bullet\right)\,\mathrm{d}y\\
    +\int_0^{+\infty}F\left(x,y,\bullet\right)f\left(t,x+y,\bullet\right)\,\mathrm{d}y\,,
\end{multline}
with $\bullet=\left(a,U,\delta,\sigma\right)$, $K$ the aggregation kernel, and $F$ the fragmentation kernel. Considering the long-term behaviour, it can be shown that $f$ takes the general form \cite{Banasiak2020a,Banasiak2020b,Sorensen1987}:
\begin{equation}
    f\left(t,x,\bullet\right)=\frac{1}{s^2\left(t,\bullet\right)}\,\varphi\left(\frac{x}{s\left(t,\bullet\right)},\bullet\right),
\end{equation}
where $s$ and $\varphi$ are two functions that depend only on time and size respectively, and such that $\int_{\mathbb{R}_+}x\varphi\left(x,\bullet\right)\,\mathrm{d}x=N$, the fixed total number of particles. The function $s$ corresponds to the average mass of the clusters according to the distribution $f$. It can be related to a size through the cluster fractal dimension $D$ by $s\propto l^D$. Then, integrating the previous expression over the size as in Eq.~\ref{eq:lbar}, one computes the steady-state average size as: 
\begin{equation}
    \overline{l}\left(t,\bullet\right)=as^\frac{1}{D}\left(t,\bullet\right)\,,
\end{equation}
with 
\begin{align}
    \lim_{t\to+\infty}s\left(t,\bullet\right)&=\left(N\frac{A\left(\bullet\right)}{B\left(\bullet\right)}\right)^\frac{1}{\alpha+2-\lambda}\\
    A\left(\bullet\right)&=\iint_{\mathbb{R}_+^2}xy\varphi\left(x,\bullet\right)\varphi\left(y,\bullet\right)K\left(x,y,\bullet\right)\,\mathrm{d}x\mathrm{d}y\\
    B\left(\bullet\right)&=\iint_{\mathbb{R}_+^2}xy\varphi\left(x+y,\bullet\right)F\left(x,y,\bullet\right)\,\mathrm{d}x\mathrm{d}y\,,
\end{align}
where $\lambda$ and $\alpha$ are the respective homogeneity coefficients of $K$ and $F$ \footnote{The homegeneity coefficients are defined as $K\left(\xi x,\xi y\right)=\xi^\lambda K\left(x,y\right)$ and $F\left(\xi x,\xi y\right)=\xi^\alpha F\left(x,y\right)$ for all $\left(x,y,\xi\right)\in\mathbb{R}_+^3$.}, assuming $\alpha+2>\lambda$. It follows from Eq.~\ref{eq:limitAverage} that
\begin{equation}
\frac{\ell}{a}=\left(N\frac{A\left(\bullet\right)}{B\left(\bullet\right)}\right)^\frac{1}{D(\alpha+2-\lambda)}.
\end{equation}
Finally, identifying with Eq. (\ref{eq:theory}), one gets 
\begin{equation}
    \label{eq:Identification}
    \frac{U}{\sigma a\delta^2}=\left( N\frac{A\left(\bullet\right)}{B\left(\bullet\right)}\right)^\frac{1+D}{D(\alpha+2-\lambda)}\,.
\end{equation}

Most of the physical quantities appear as multiplicative factors, in the sense that, for example, the coagulation kernel $K$ is usually built with a factor $U/\eta$ with $\eta$ the viscosity of the fluid and no other dependence on physical quantities besides the variables $x$ and $y$. Therefore, the ratio $NA\left(\bullet\right)/B\left(\bullet\right)$ can be considered as the product of a dimensionless number built in a similar manner as the adhesion number and another factor that depends only on the shape of respective kernels without involving any additional physical parameter. Therefore, in order to keep Eq.~(\ref{eq:Identification}) true in general, one should impose that the exponent is 1, which leads to

\begin{equation}
    1=D\left(\alpha+1-\lambda\right)\,.
\label{eq:D}
\end{equation}
This is an important result which, to our knowledge, has not been reported in the literature before. Indeed, Eq.~(\ref{eq:D}) allows one to relate the first mechanical approach to  population balance models, where the homogeneity coefficients may seem disconnected from physical and measurable quantities. This result also emphasizes the fact that the adhesion number is a correct measure of the relative importance of aggregation versus disaggregation.

We note that aggregation kernels are pretty well covered, either in terms of theoretical solutions \cite{Banasiak2020a,Banasiak2020b,Wattis2006,Spicer1996} with simple sums or products, or in terms of a physical construction, e.g., based on collisions, thermal fluctuations, and diffusion \cite{Kryven2014,Barthelmes2003}. Fragmentation kernels, however, are more poorly controlled, in the sense that some theoretical solutions impose some strong conditions on these kernels without much physical justification \cite{Banasiak2020a,Banasiak2020b,Lattuada2016,Puisto2012}, so that the expressions of the fragmentation kernels remain mostly empirical or semi-empirical \cite{Delichatsios1976,Kuster1991}. Therefore, although the use of coagulation-fragmentation equations is well established, relating some of the main terms to physical phenomena, such as the interaction potential and the flow stresses, is an important step yet to be fully achieved.

\subsection{Comparison with simulations and experiments}

Relating the model proposed in Sect.~\ref{sec:Model} to the simulations of Sect.~\ref{sec:Numerical}, we expect the exponents in Table~\ref{tab:ExponentsFactor2} to be linked to the fractal dimension respectively by $1/\left(1+D\right)$ for the steady-state size $\ell$, and  $D/\left(1+D\right)$ for the equilibrium mass $k_{\min}$. The broad variability of the exponents does not allow to properly extract a fractal dimension from the simulations. Nevertheless, as discussed above in Sect.~\ref{sec:SimRes}, the shear-induced clusters are almost dense, so that we may assume a fractal dimension $D$ close to 2. Such a compactness most probably results from the specific interaction potential, i.e. a 12-6 Lennard-Jones potential, which is a central-force potential and is likely to lead to clusters with droplet-like shapes. We note that $D=2$ would be consistent with the exponent $\beta=\left(1+D\right)^{-1}=1/3$ expected for $\ell$ as a function of $\mathrm{Ad}$, at least in the smaller geometry, since Table~\ref{tab:ExponentsFactor2} reports exponents of 0.35 and 0.27 for $h=103$. 

\begin{figure}
\begin{centering}
\includegraphics[width=\columnwidth]{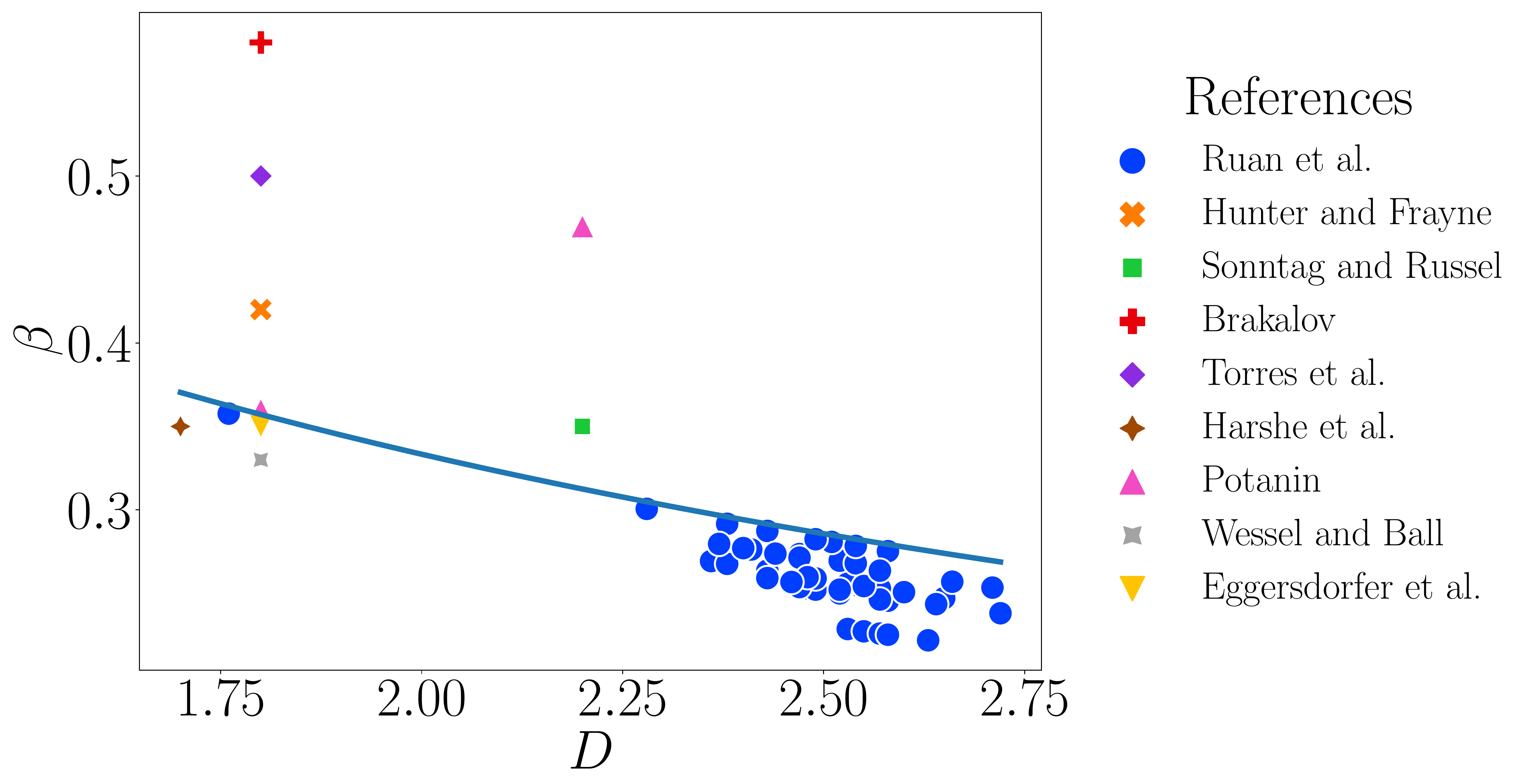}
\par\end{centering}
\caption{Power-law exponent $\beta$ in $\ell/a=\mathrm{Ad}^\beta$ as a function of
the fractal dimension $D$ of the aggregates extracted from Refs.~\cite{Ruan2020,Hunter1980,Sonntag1986,Sonntag1987,Sonntag1987b,Brakalov1987,Torres1991a,Torres1991b,Harshe2011,Potanin1991,Potanin1992,Potanin1996,Wessel1992,Higashitani1998,Higashitani2001,Eggersdorfer2010,Kimbonguila2014} ($\bullet$) and compared to the model prediction $\beta=\left(1+D\right)^{-1}$ (solid line). 
\label{fig:Exponent}}
\end{figure}

Moreover, the results may be compared to the recent three-dimensional simulations under simple shear flow by Ruan \textit{et al.} \cite{Ruan2020}, which also report rather dense shear-induced clusters at steady state. Focusing both on the cluster size and on the average number of particles in a cluster as in the present work, the authors extract the cluster fractal dimension $D$ and show that $\ell/a$ scales as a power-law of the shear stress. The exponents $\beta$ for $\ell$ reported in Ref.~\cite{Ruan2020} are replotted as a function of $D$ as blue circles in Fig.~\ref{fig:Exponent} and show good agreement with the prediction $\beta=\left(1+D\right)^{-1}$. Note that this is also consistent with the empirical expression proposed in Ref.~\cite{Ruan2020}, namely $\beta=S/(S+\langle D\rangle)$, where $\langle D\rangle$ denotes the average cluster fractal dimension, and the fitting parameter $S$ is found to be close to 1.

On the experimental side, a number of works have reported results assessing the size of aggregates under the application of a mechanical stress. However, only a handful of papers explicitly state the values of the various parameters $a$, $U$, $\sigma$, $\delta$, and $D$. Table \ref{tab:LiteratureResults} and Fig. \ref{fig:Exponent} gather a selection of such previous works. First, Table~\ref{tab:LiteratureResults} shows that Eq.~(\ref{eq:theory}) predicts a typical cluster size $\ell_\mathrm{th}$ which is of the same order as the experimental size $\ell_\mathrm{exp}$. All the parameters fall into commonly known ranges and justify that the previous approach may be a good proxy to evaluate the most stable cluster size.
Second,
Fig. \ref{fig:Exponent} probes the sensitivity of $\ell/a$ with the adhesion number $\mathrm{Ad}$ by plotting the power-law exponent $\beta$ in $\ell/a=\mathrm{Ad}^\beta$ as a function of $D$. Except for a few points that lie far from the average estimation,
most exponents are gathered along the prediction of Eq. (\ref{eq:theory}), consistently with the numerical results of Ref.~\cite{Ruan2020}, which provides strong support for the approach described in Sect.~\ref{sec:Model}.

\begin{table}
    \centering
    \begin{tabular}{ccc}
        \hline
        Source & \cite{Gibaud2020a,Gibaud2020b,Dages2021,Varga2019} & \cite{Nguyen2011,Klimchitskaya2000,Visser1972,Waite2001} \\
        \hline
        $\ell_\mathrm{exp}$ (nm) & 500 & 60\\
        $a$ (nm)& 150 & 6.5\\
        $U/k_BT$ & 20 & 60\\
        $\sigma$ (MPa) & 0.1 & 2\\
        $\delta\,\left(\mathring{\mathrm{A}}\right)$  & 7 & 2\\
        $D$ & 2.6 & 1.88\\
        \hline
        $\ell_\mathrm{th}$ (nm) & 300 & 55\\
        \hline
    \end{tabular}
    \caption{Experimental parameters and measurements of the aggregate size from the literature. $\ell_\mathrm{exp}$ is the direct measurement of the aggregate size, while $\ell_\mathrm{th}$ is the aggregate size estimated using Eq.~(\ref{eq:theory}) based on the parameters defined in the text and reported in the references of the first column.}
    \label{tab:LiteratureResults}
\end{table}

\begin{figure}
    \centering
    \includegraphics[width=\columnwidth]{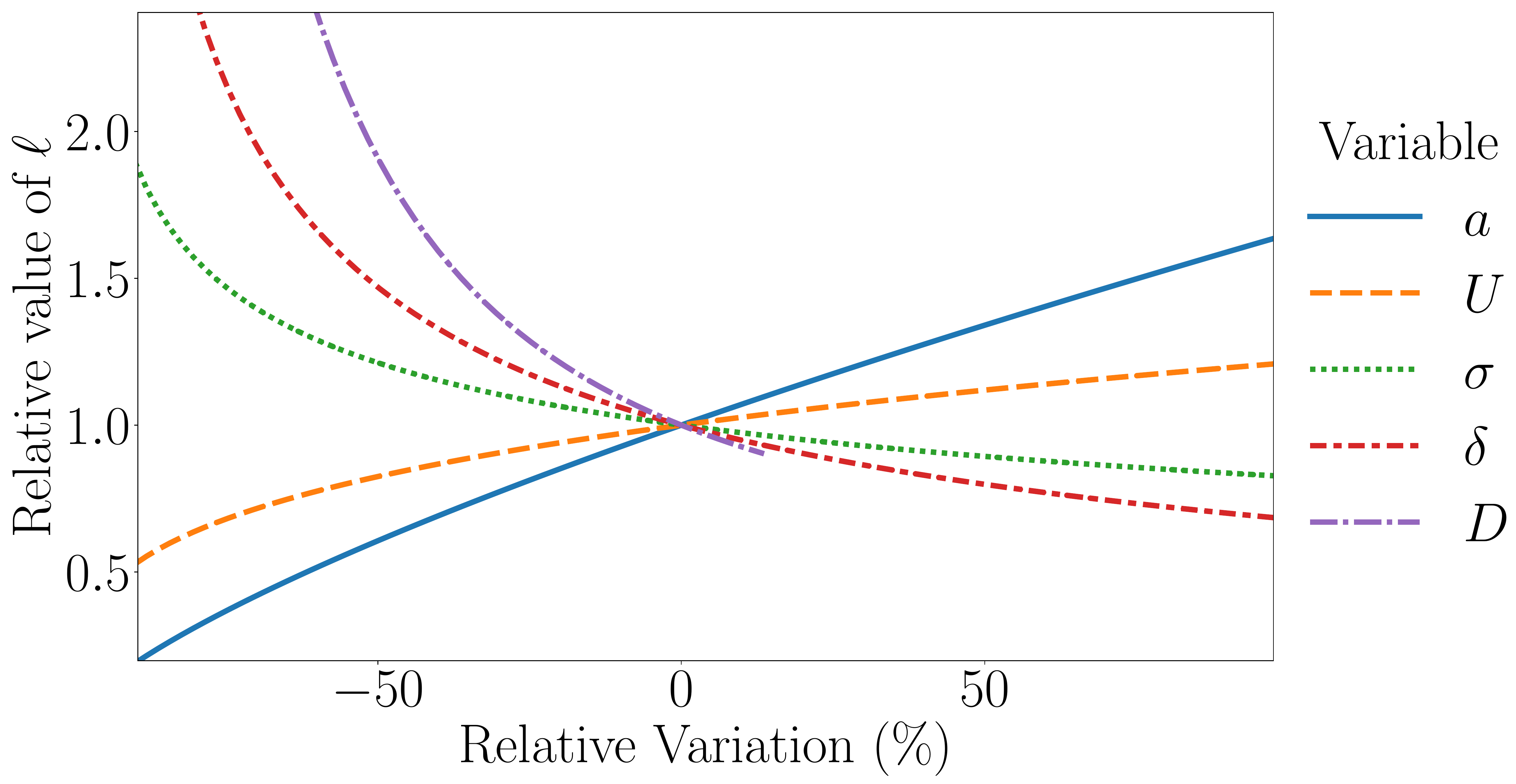}
    \caption{Sensitivity study of $\ell$ predicted by Eq.~(\ref{eq:theory}) upon relative variations of the different control parameters around the values listed in the first line of Table~\ref{tab:LiteratureResults} except for the fractal dimension, which is varied from 1 to 3.}
    \label{fig:Sensitivityell}
\end{figure}

Finally, Fig.~\ref{fig:Sensitivityell} provides a sensitivity study of the value of $\ell$ upon the different parameters of Eq. (\ref{eq:theory}) based on Table~\ref{tab:LiteratureResults}. It is clear that the most critical parameters are $D$ and $\delta$ as expected from Eq.~(\ref{eq:theory}). Therefore, particular attention must be considered to assess accurately these parameters. Yet, when one of these two parameters are unknown, the model can be used to estimate $D$ or $\delta$ with good accuracy based on measurements of $\ell$.

\section{Conclusion}
\label{sec:Conclusion}

We have investigated simple two-dimensional numerical simulations and proposed a theoretical description of aggregation and disaggregation processes in colloidal dispersions under shear. We have shown that the classically reported power-law increase of the average cluster size with the adhesion number can be rationalized through a dynamical balance of energy between aggregation and disaggregation. 
In particular, the algebraic dependence of the power-law exponent with the cluster fractal dimension predicted by the approach appears to be in good agreement with simulations and previous experiments. Still, one should remain careful about a direct inference of one simple general formula as provided here, since many additional parameters may be involved in the size selection, including time through aging phenomena, spatial confinement, and individual cluster dynamics.

\acknowledgments
The authors wish to thank Emanuela Del Gado, Thibaut Divoux, and Thomas Gibaud for fruitful discussions.

\bibliography{Bibliographie_these}

\begin{thebibliography}{96}%
\makeatletter
\providecommand \@ifxundefined [1]{%
 \@ifx{#1\undefined}
}%
\providecommand \@ifnum [1]{%
 \ifnum #1\expandafter \@firstoftwo
 \else \expandafter \@secondoftwo
 \fi
}%
\providecommand \@ifx [1]{%
 \ifx #1\expandafter \@firstoftwo
 \else \expandafter \@secondoftwo
 \fi
}%
\providecommand \natexlab [1]{#1}%
\providecommand \enquote  [1]{``#1''}%
\providecommand \bibnamefont  [1]{#1}%
\providecommand \bibfnamefont [1]{#1}%
\providecommand \citenamefont [1]{#1}%
\providecommand \href@noop [0]{\@secondoftwo}%
\providecommand \href [0]{\begingroup \@sanitize@url \@href}%
\providecommand \@href[1]{\@@startlink{#1}\@@href}%
\providecommand \@@href[1]{\endgroup#1\@@endlink}%
\providecommand \@sanitize@url [0]{\catcode `\\12\catcode `\$12\catcode
  `\&12\catcode `\#12\catcode `\^12\catcode `\_12\catcode `\%12\relax}%
\providecommand \@@startlink[1]{}%
\providecommand \@@endlink[0]{}%
\providecommand \url  [0]{\begingroup\@sanitize@url \@url }%
\providecommand \@url [1]{\endgroup\@href {#1}{\urlprefix }}%
\providecommand \urlprefix  [0]{URL }%
\providecommand \Eprint [0]{\href }%
\providecommand \doibase [0]{https://doi.org/}%
\providecommand \selectlanguage [0]{\@gobble}%
\providecommand \bibinfo  [0]{\@secondoftwo}%
\providecommand \bibfield  [0]{\@secondoftwo}%
\providecommand \translation [1]{[#1]}%
\providecommand \BibitemOpen [0]{}%
\providecommand \bibitemStop [0]{}%
\providecommand \bibitemNoStop [0]{.\EOS\space}%
\providecommand \EOS [0]{\spacefactor3000\relax}%
\providecommand \BibitemShut  [1]{\csname bibitem#1\endcsname}%
\let\auto@bib@innerbib\@empty
\bibitem [{\citenamefont {Macosko}(1994)}]{Macosko1994}%
  \BibitemOpen
  \bibfield  {author} {\bibinfo {author} {\bibfnamefont {C.~W.}\ \bibnamefont
  {Macosko}},\ }\href@noop {} {\emph {\bibinfo {title} {{Rheology : Principles,
  Measurements and Applications}}}},\ edited by\ \bibinfo {editor}
  {\bibfnamefont {C.~W.}\ \bibnamefont {Macosko}}\ (\bibinfo  {publisher} {John
  Wiley and Sons},\ \bibinfo {year} {1994})\BibitemShut {NoStop}%
\bibitem [{\citenamefont {Mewis}\ and\ \citenamefont
  {Wagner}(2012)}]{Mewis2012}%
  \BibitemOpen
  \bibfield  {author} {\bibinfo {author} {\bibfnamefont {J.}~\bibnamefont
  {Mewis}}\ and\ \bibinfo {author} {\bibfnamefont {N.~J.}\ \bibnamefont
  {Wagner}},\ }\href {https://doi.org/10.1017/CBO9780511977978} {\emph
  {\bibinfo {title} {{Colloidal Suspension Rheology}}}},\ edited by\ \bibinfo
  {editor} {\bibfnamefont {J.}~\bibnamefont {Mewis}}\ and\ \bibinfo {editor}
  {\bibfnamefont {N.~J.}\ \bibnamefont {Wagner}}\ (\bibinfo  {publisher}
  {Cambridge Univesity Press},\ \bibinfo {year} {2012})\BibitemShut {NoStop}%
\bibitem [{\citenamefont {Wagner}\ and\ \citenamefont
  {Mewis}(2021)}]{Wagner2021}%
  \BibitemOpen
  \bibfield  {author} {\bibinfo {author} {\bibfnamefont {N.~J.}\ \bibnamefont
  {Wagner}}\ and\ \bibinfo {author} {\bibfnamefont {J.}~\bibnamefont {Mewis}},\
  }\href {https://doi.org/10.1017/9781108394826} {\emph {\bibinfo {title}
  {{Theory and Applications of Colloidal Suspension Rheology}}}},\ edited by\
  \bibinfo {editor} {\bibfnamefont {N.~J.}\ \bibnamefont {Wagner}}\ and\
  \bibinfo {editor} {\bibfnamefont {J.}~\bibnamefont {Mewis}}\ (\bibinfo
  {publisher} {Cambridge University Press},\ \bibinfo {year}
  {2021})\BibitemShut {NoStop}%
\bibitem [{\citenamefont {Hengl}\ \emph {et~al.}(2014)\citenamefont {Hengl},
  \citenamefont {Jin}, \citenamefont {Pignon}, \citenamefont {Baup},
  \citenamefont {Mollard}, \citenamefont {Gondrexon}, \citenamefont {Magnin},
  \citenamefont {Michot},\ and\ \citenamefont {Paineau}}]{Hengl2014}%
  \BibitemOpen
  \bibfield  {author} {\bibinfo {author} {\bibfnamefont {N.}~\bibnamefont
  {Hengl}}, \bibinfo {author} {\bibfnamefont {Y.}~\bibnamefont {Jin}}, \bibinfo
  {author} {\bibfnamefont {F.}~\bibnamefont {Pignon}}, \bibinfo {author}
  {\bibfnamefont {S.}~\bibnamefont {Baup}}, \bibinfo {author} {\bibfnamefont
  {R.}~\bibnamefont {Mollard}}, \bibinfo {author} {\bibfnamefont
  {N.}~\bibnamefont {Gondrexon}}, \bibinfo {author} {\bibfnamefont
  {A.}~\bibnamefont {Magnin}}, \bibinfo {author} {\bibfnamefont
  {L.}~\bibnamefont {Michot}},\ and\ \bibinfo {author} {\bibfnamefont
  {E.}~\bibnamefont {Paineau}},\ }\bibfield  {title} {\bibinfo {title} {{A new
  way to apply ultrasound in cross-flow ultrafiltration: Application to
  colloidal suspensions}},\ }\href
  {https://doi.org/10.1016/j.ultsonch.2013.11.008} {\bibfield  {journal}
  {\bibinfo  {journal} {Ultrasonics Sonochemistry}\ }\textbf {\bibinfo {volume}
  {21}},\ \bibinfo {pages} {1018} (\bibinfo {year} {2014})}\BibitemShut
  {NoStop}%
\bibitem [{\citenamefont {Ioannidou}\ \emph {et~al.}(2016)\citenamefont
  {Ioannidou}, \citenamefont {Krakowiak}, \citenamefont {Bauchy}, \citenamefont
  {Hoover}, \citenamefont {Masoero}, \citenamefont {Yip}, \citenamefont {Ulm},
  \citenamefont {Levitz}, \citenamefont {Pellenq},\ and\ \citenamefont
  {Del~Gado}}]{Ioannidou2016}%
  \BibitemOpen
  \bibfield  {author} {\bibinfo {author} {\bibfnamefont {K.}~\bibnamefont
  {Ioannidou}}, \bibinfo {author} {\bibfnamefont {K.~J.}\ \bibnamefont
  {Krakowiak}}, \bibinfo {author} {\bibfnamefont {M.}~\bibnamefont {Bauchy}},
  \bibinfo {author} {\bibfnamefont {C.~G.}\ \bibnamefont {Hoover}}, \bibinfo
  {author} {\bibfnamefont {E.}~\bibnamefont {Masoero}}, \bibinfo {author}
  {\bibfnamefont {S.}~\bibnamefont {Yip}}, \bibinfo {author} {\bibfnamefont
  {F.-J.}\ \bibnamefont {Ulm}}, \bibinfo {author} {\bibfnamefont
  {P.}~\bibnamefont {Levitz}}, \bibinfo {author} {\bibfnamefont {R.~J.-M.}\
  \bibnamefont {Pellenq}},\ and\ \bibinfo {author} {\bibfnamefont
  {E.}~\bibnamefont {Del~Gado}},\ }\bibfield  {title} {\bibinfo {title}
  {Mesoscale texture of cement hydrates},\ }\href
  {https://doi.org/10.1073/pnas.1520487113} {\bibfield  {journal} {\bibinfo
  {journal} {Proceedings of the National Academy of Sciences of the United
  States of America}\ }\textbf {\bibinfo {volume} {113}},\ \bibinfo {pages}
  {2029} (\bibinfo {year} {2016})}\BibitemShut {NoStop}%
\bibitem [{\citenamefont {Awad}\ \emph {et~al.}(2012)\citenamefont {Awad},
  \citenamefont {Moharram}, \citenamefont {Shaltout},\ and\ \citenamefont
  {Youssef}}]{Awad2012}%
  \BibitemOpen
  \bibfield  {author} {\bibinfo {author} {\bibfnamefont {T.~S.}\ \bibnamefont
  {Awad}}, \bibinfo {author} {\bibfnamefont {H.~A.}\ \bibnamefont {Moharram}},
  \bibinfo {author} {\bibfnamefont {O.~E.}\ \bibnamefont {Shaltout}},\ and\
  \bibinfo {author} {\bibfnamefont {M.~M.}\ \bibnamefont {Youssef}},\
  }\bibfield  {title} {\bibinfo {title} {{Applications of ultrasound in
  analysis, processing and quality control of food: A review}},\ }\href
  {https://doi.org/10.1016/j.foodres.2012.05.004} {\bibfield  {journal}
  {\bibinfo  {journal} {Food Research International}\ }\textbf {\bibinfo
  {volume} {48}},\ \bibinfo {pages} {410} (\bibinfo {year} {2012})}\BibitemShut
  {NoStop}%
\bibitem [{\citenamefont {Knorr}\ \emph {et~al.}(2004)\citenamefont {Knorr},
  \citenamefont {Zenker}, \citenamefont {Heinz},\ and\ \citenamefont
  {Lee}}]{Knorr2004}%
  \BibitemOpen
  \bibfield  {author} {\bibinfo {author} {\bibfnamefont {D.}~\bibnamefont
  {Knorr}}, \bibinfo {author} {\bibfnamefont {M.}~\bibnamefont {Zenker}},
  \bibinfo {author} {\bibfnamefont {V.}~\bibnamefont {Heinz}},\ and\ \bibinfo
  {author} {\bibfnamefont {D.-U.}\ \bibnamefont {Lee}},\ }\bibfield  {title}
  {\bibinfo {title} {Applications and potential of ultrasonics in food
  processing},\ }\href {https://doi.org/10.1016/j.tifs.2003.12.00} {\bibfield
  {journal} {\bibinfo  {journal} {Trends in Food Science \& Technology}\
  }\textbf {\bibinfo {volume} {15}},\ \bibinfo {pages} {261} (\bibinfo {year}
  {2004})}\BibitemShut {NoStop}%
\bibitem [{\citenamefont {Chandrapal}\ \emph {et~al.}(2012)\citenamefont
  {Chandrapal}, \citenamefont {Oliver}, \citenamefont {Kentish},\ and\
  \citenamefont {Ashokkumar}}]{Chandrapala2012}%
  \BibitemOpen
  \bibfield  {author} {\bibinfo {author} {\bibfnamefont {J.}~\bibnamefont
  {Chandrapal}}, \bibinfo {author} {\bibfnamefont {C.}~\bibnamefont {Oliver}},
  \bibinfo {author} {\bibfnamefont {S.}~\bibnamefont {Kentish}},\ and\ \bibinfo
  {author} {\bibfnamefont {M.}~\bibnamefont {Ashokkumar}},\ }\bibfield  {title}
  {\bibinfo {title} {Ultrasonics in food processing},\ }\href
  {https://doi.org/10.1016/j.ultsonch.2012.01.010} {\bibfield  {journal}
  {\bibinfo  {journal} {Ultrasonics Sonochemistry}\ }\textbf {\bibinfo {volume}
  {19}},\ \bibinfo {pages} {975} (\bibinfo {year} {2012})}\BibitemShut
  {NoStop}%
\bibitem [{\citenamefont {Bonn}\ \emph {et~al.}(2017)\citenamefont {Bonn},
  \citenamefont {Denn}, \citenamefont {Berthier}, \citenamefont {Divoux},\ and\
  \citenamefont {Manneville}}]{Bonn2017}%
  \BibitemOpen
  \bibfield  {author} {\bibinfo {author} {\bibfnamefont {D.}~\bibnamefont
  {Bonn}}, \bibinfo {author} {\bibfnamefont {M.~M.}\ \bibnamefont {Denn}},
  \bibinfo {author} {\bibfnamefont {L.}~\bibnamefont {Berthier}}, \bibinfo
  {author} {\bibfnamefont {T.}~\bibnamefont {Divoux}},\ and\ \bibinfo {author}
  {\bibfnamefont {S.}~\bibnamefont {Manneville}},\ }\bibfield  {title}
  {\bibinfo {title} {Yield stress materials in soft condensed matter},\ }\href
  {https://doi.org/10.1103/RevModPhys.89.035005} {\bibfield  {journal}
  {\bibinfo  {journal} {Reviews of Modern Physics}\ }\textbf {\bibinfo {volume}
  {89}},\ \bibinfo {pages} {1} (\bibinfo {year} {2017})},\ \bibinfo {note}
  {035005}\BibitemShut {NoStop}%
\bibitem [{\citenamefont {Vassileva}\ \emph {et~al.}(2005)\citenamefont
  {Vassileva}, \citenamefont {van~den Ende}, \citenamefont {Mugele},\ and\
  \citenamefont {Mellema}}]{Vassileva2005}%
  \BibitemOpen
  \bibfield  {author} {\bibinfo {author} {\bibfnamefont {N.~D.}\ \bibnamefont
  {Vassileva}}, \bibinfo {author} {\bibfnamefont {D.}~\bibnamefont {van~den
  Ende}}, \bibinfo {author} {\bibfnamefont {F.}~\bibnamefont {Mugele}},\ and\
  \bibinfo {author} {\bibfnamefont {J.}~\bibnamefont {Mellema}},\ }\bibfield
  {title} {\bibinfo {title} {{Capillary Forces between spherical Particles
  Floating at a Liquid-Liquid Interface}},\ }\href
  {https://doi.org/10.1021/la051186o} {\bibfield  {journal} {\bibinfo
  {journal} {Langmuir}\ }\textbf {\bibinfo {volume} {21}},\ \bibinfo {pages}
  {11190} (\bibinfo {year} {2005})}\BibitemShut {NoStop}%
\bibitem [{\citenamefont {Wessel}\ and\ \citenamefont
  {Ball}(1992)}]{Wessel1992}%
  \BibitemOpen
  \bibfield  {author} {\bibinfo {author} {\bibfnamefont {R.}~\bibnamefont
  {Wessel}}\ and\ \bibinfo {author} {\bibfnamefont {R.~C.}\ \bibnamefont
  {Ball}},\ }\bibfield  {title} {\bibinfo {title} {Fractal aggregates and gels
  in shear flow},\ }\href {https://doi.org/10.1103/PhysRevA.46.R3008}
  {\bibfield  {journal} {\bibinfo  {journal} {Physical Review A}\ }\textbf
  {\bibinfo {volume} {46}},\ \bibinfo {pages} {R3008} (\bibinfo {year}
  {1992})}\BibitemShut {NoStop}%
\bibitem [{\citenamefont {Shih}\ \emph {et~al.}(1990)\citenamefont {Shih},
  \citenamefont {Shih}, \citenamefont {Kim}, \citenamefont {Liu},\ and\
  \citenamefont {Aksay}}]{Shih1990}%
  \BibitemOpen
  \bibfield  {author} {\bibinfo {author} {\bibfnamefont {W.-H.}\ \bibnamefont
  {Shih}}, \bibinfo {author} {\bibfnamefont {W.~Y.}\ \bibnamefont {Shih}},
  \bibinfo {author} {\bibfnamefont {S.-I.}\ \bibnamefont {Kim}}, \bibinfo
  {author} {\bibfnamefont {J.}~\bibnamefont {Liu}},\ and\ \bibinfo {author}
  {\bibfnamefont {I.~A.}\ \bibnamefont {Aksay}},\ }\bibfield  {title} {\bibinfo
  {title} {Scaling behavior of the elastic properties of colloidal gels},\
  }\href {https://doi.org/10.1103/PhysRevA.42.4772} {\bibfield  {journal}
  {\bibinfo  {journal} {Physical Review A}\ }\textbf {\bibinfo {volume} {42}},\
  \bibinfo {pages} {4772} (\bibinfo {year} {1990})}\BibitemShut {NoStop}%
\bibitem [{\citenamefont {Sorensen}(2001)}]{Sorensen2001}%
  \BibitemOpen
  \bibfield  {author} {\bibinfo {author} {\bibfnamefont {C.~M.}\ \bibnamefont
  {Sorensen}},\ }\bibfield  {title} {\bibinfo {title} {{Light Scattering by
  Fractal Aggregates: A Review}},\ }\href
  {https://doi.org/10.1080/02786820117868} {\bibfield  {journal} {\bibinfo
  {journal} {Aerosol Science Technology}\ }\textbf {\bibinfo {volume} {35}},\
  \bibinfo {pages} {648} (\bibinfo {year} {2001})}\BibitemShut {NoStop}%
\bibitem [{\citenamefont {Lin}\ \emph {et~al.}(1990)\citenamefont {Lin},
  \citenamefont {Klein}, \citenamefont {Lindsay}, \citenamefont {Weitz},
  \citenamefont {Ball},\ and\ \citenamefont {Meakin}}]{Lin1990a}%
  \BibitemOpen
  \bibfield  {author} {\bibinfo {author} {\bibfnamefont {M.~Y.}\ \bibnamefont
  {Lin}}, \bibinfo {author} {\bibfnamefont {R.}~\bibnamefont {Klein}}, \bibinfo
  {author} {\bibfnamefont {H.~M.}\ \bibnamefont {Lindsay}}, \bibinfo {author}
  {\bibfnamefont {D.~A.}\ \bibnamefont {Weitz}}, \bibinfo {author}
  {\bibfnamefont {R.~C.}\ \bibnamefont {Ball}},\ and\ \bibinfo {author}
  {\bibfnamefont {P.}~\bibnamefont {Meakin}},\ }\bibfield  {title} {\bibinfo
  {title} {{The Structure of Fractal Colloidal Aggregates of Finite Extent}},\
  }\href {https://doi.org/10.1016/0021-9797(90)90061-R} {\bibfield  {journal}
  {\bibinfo  {journal} {Journal of Colloid and Interface Science}\ }\textbf
  {\bibinfo {volume} {137}},\ \bibinfo {pages} {263} (\bibinfo {year}
  {1990})}\BibitemShut {NoStop}%
\bibitem [{\citenamefont {Lazzari}\ \emph {et~al.}(2016)\citenamefont
  {Lazzari}, \citenamefont {Nicoud}, \citenamefont {Jaquet}, \citenamefont
  {Lattuada},\ and\ \citenamefont {Morbidelli}}]{Lazzari2016}%
  \BibitemOpen
  \bibfield  {author} {\bibinfo {author} {\bibfnamefont {S.}~\bibnamefont
  {Lazzari}}, \bibinfo {author} {\bibfnamefont {L.}~\bibnamefont {Nicoud}},
  \bibinfo {author} {\bibfnamefont {B.}~\bibnamefont {Jaquet}}, \bibinfo
  {author} {\bibfnamefont {M.}~\bibnamefont {Lattuada}},\ and\ \bibinfo
  {author} {\bibfnamefont {M.}~\bibnamefont {Morbidelli}},\ }\bibfield  {title}
  {\bibinfo {title} {Fractal-like structures in colloid science},\ }\href
  {https://doi.org/10.1016/j.cis.2016.05.002} {\bibfield  {journal} {\bibinfo
  {journal} {Advances in Colloid and Interface Science}\ }\textbf {\bibinfo
  {volume} {235}},\ \bibinfo {pages} {1} (\bibinfo {year} {2016})}\BibitemShut
  {NoStop}%
\bibitem [{\citenamefont {Grassberger}(1985)}]{Grassberger1985}%
  \BibitemOpen
  \bibfield  {author} {\bibinfo {author} {\bibfnamefont {P.}~\bibnamefont
  {Grassberger}},\ }\bibfield  {title} {\bibinfo {title} {On the spreading of
  two-dimensional percolation},\ }\href
  {https://doi.org/10.1088/0305-4470/18/4/005} {\bibfield  {journal} {\bibinfo
  {journal} {Journal of Physics A: Mathematical and General}\ }\textbf
  {\bibinfo {volume} {18}},\ \bibinfo {pages} {L215} (\bibinfo {year}
  {1985})}\BibitemShut {NoStop}%
\bibitem [{\citenamefont {Bantawa}\ \emph {et~al.}(2021)\citenamefont
  {Bantawa}, \citenamefont {Fontaine-Seiler}, \citenamefont {Olmsted},\ and\
  \citenamefont {Del~Gado}}]{Bantawa2021}%
  \BibitemOpen
  \bibfield  {author} {\bibinfo {author} {\bibfnamefont {M.}~\bibnamefont
  {Bantawa}}, \bibinfo {author} {\bibfnamefont {W.~A.}\ \bibnamefont
  {Fontaine-Seiler}}, \bibinfo {author} {\bibfnamefont {P.~D.}\ \bibnamefont
  {Olmsted}},\ and\ \bibinfo {author} {\bibfnamefont {E.}~\bibnamefont
  {Del~Gado}},\ }\bibfield  {title} {\bibinfo {title} {Microscopic interactions
  and emerging elasticity in model soft particulate gels},\ }\href
  {https://doi.org/10.1088/1361-648X/ac14f6} {\bibfield  {journal} {\bibinfo
  {journal} {Journal of Physics: Condensed Matter}\ }\textbf {\bibinfo {volume}
  {33}},\ \bibinfo {pages} {1} (\bibinfo {year} {2021})}\BibitemShut {NoStop}%
\bibitem [{\citenamefont {Weitz}\ and\ \citenamefont
  {Oliveria}(1984)}]{Weitz1984}%
  \BibitemOpen
  \bibfield  {author} {\bibinfo {author} {\bibfnamefont {D.~A.}\ \bibnamefont
  {Weitz}}\ and\ \bibinfo {author} {\bibfnamefont {M.}~\bibnamefont
  {Oliveria}},\ }\bibfield  {title} {\bibinfo {title} {{Fractal Structures
  Formed by Kinetic Aggregation of Aqueous Gold Colloids}},\ }\href
  {https://doi.org/10.1103/PhysRevLett.52.1433} {\bibfield  {journal} {\bibinfo
   {journal} {Physical Review Letters}\ }\textbf {\bibinfo {volume} {52}},\
  \bibinfo {pages} {1433} (\bibinfo {year} {1984})}\BibitemShut {NoStop}%
\bibitem [{\citenamefont {Weitz}\ \emph {et~al.}(1985)\citenamefont {Weitz},
  \citenamefont {Huang}, \citenamefont {Lin},\ and\ \citenamefont
  {Sung}}]{Weitz1985}%
  \BibitemOpen
  \bibfield  {author} {\bibinfo {author} {\bibfnamefont {D.~A.}\ \bibnamefont
  {Weitz}}, \bibinfo {author} {\bibfnamefont {J.~S.}\ \bibnamefont {Huang}},
  \bibinfo {author} {\bibfnamefont {M.~Y.}\ \bibnamefont {Lin}},\ and\ \bibinfo
  {author} {\bibfnamefont {J.}~\bibnamefont {Sung}},\ }\bibfield  {title}
  {\bibinfo {title} {{Limits of the Fractal Dimension for Irreversible Kinetic
  Aggregation of Gold Colloids}},\ }\href
  {https://doi.org/10.1103/PhysRevLett.54.1416} {\bibfield  {journal} {\bibinfo
   {journal} {Physical Review Letters}\ }\textbf {\bibinfo {volume} {54}},\
  \bibinfo {pages} {1416} (\bibinfo {year} {1985})}\BibitemShut {NoStop}%
\bibitem [{\citenamefont {Hoekstra}\ \emph {et~al.}(2003)\citenamefont
  {Hoekstra}, \citenamefont {Vermant}, \citenamefont {Mewis},\ and\
  \citenamefont {Fuller}}]{Hoekstra:2003}%
  \BibitemOpen
  \bibfield  {author} {\bibinfo {author} {\bibfnamefont {H.}~\bibnamefont
  {Hoekstra}}, \bibinfo {author} {\bibfnamefont {J.}~\bibnamefont {Vermant}},
  \bibinfo {author} {\bibfnamefont {J.}~\bibnamefont {Mewis}},\ and\ \bibinfo
  {author} {\bibfnamefont {G.}~\bibnamefont {Fuller}},\ }\bibfield  {title}
  {\bibinfo {title} {Flow-induced anisotropy and reversible aggregation in
  two-dimensional suspensions},\ }\href@noop {} {\bibfield  {journal} {\bibinfo
   {journal} {Langmuir}\ }\textbf {\bibinfo {volume} {19}},\ \bibinfo {pages}
  {9134} (\bibinfo {year} {2003})}\BibitemShut {NoStop}%
\bibitem [{\citenamefont {Nguyen}\ \emph {et~al.}(2011)\citenamefont {Nguyen},
  \citenamefont {Rouxel}, \citenamefont {Hadji}, \citenamefont {Vincent},\ and\
  \citenamefont {Fort}}]{Nguyen2011}%
  \BibitemOpen
  \bibfield  {author} {\bibinfo {author} {\bibfnamefont {V.~S.}\ \bibnamefont
  {Nguyen}}, \bibinfo {author} {\bibfnamefont {D.}~\bibnamefont {Rouxel}},
  \bibinfo {author} {\bibfnamefont {R.}~\bibnamefont {Hadji}}, \bibinfo
  {author} {\bibfnamefont {B.}~\bibnamefont {Vincent}},\ and\ \bibinfo {author}
  {\bibfnamefont {Y.}~\bibnamefont {Fort}},\ }\bibfield  {title} {\bibinfo
  {title} {{Effect of ultrasonication and dispersion stability on the cluster
  size of alumina nanoscale particles in aqueous solutions}},\ }\href
  {https://doi.org/10.1016/j.ultsonch.2010.07.003} {\bibfield  {journal}
  {\bibinfo  {journal} {Ultrasonics Sonochemistry}\ }\textbf {\bibinfo {volume}
  {18}},\ \bibinfo {pages} {382} (\bibinfo {year} {2011})}\BibitemShut
  {NoStop}%
\bibitem [{\citenamefont {Richards}\ \emph {et~al.}(2017)\citenamefont
  {Richards}, \citenamefont {Hipp}, \citenamefont {Riley}, \citenamefont
  {Wagner},\ and\ \citenamefont {Butler}}]{Richards2017}%
  \BibitemOpen
  \bibfield  {author} {\bibinfo {author} {\bibfnamefont {J.~J.}\ \bibnamefont
  {Richards}}, \bibinfo {author} {\bibfnamefont {J.~B.}\ \bibnamefont {Hipp}},
  \bibinfo {author} {\bibfnamefont {J.~K.}\ \bibnamefont {Riley}}, \bibinfo
  {author} {\bibfnamefont {N.~J.}\ \bibnamefont {Wagner}},\ and\ \bibinfo
  {author} {\bibfnamefont {P.~D.}\ \bibnamefont {Butler}},\ }\bibfield  {title}
  {\bibinfo {title} {{Clustering and Percolation in Suspensions of Carbon
  Black}},\ }\href {https://doi.org/10.1021/acs.langmuir.7b02538} {\bibfield
  {journal} {\bibinfo  {journal} {Langmuir}\ }\textbf {\bibinfo {volume}
  {33}},\ \bibinfo {pages} {12260} (\bibinfo {year} {2017})}\BibitemShut
  {NoStop}%
\bibitem [{\citenamefont {Masschaele}\ \emph {et~al.}(2009)\citenamefont
  {Masschaele}, \citenamefont {Fransaer},\ and\ \citenamefont
  {Vermant}}]{Masschaele2009}%
  \BibitemOpen
  \bibfield  {author} {\bibinfo {author} {\bibfnamefont {K.}~\bibnamefont
  {Masschaele}}, \bibinfo {author} {\bibfnamefont {J.}~\bibnamefont
  {Fransaer}},\ and\ \bibinfo {author} {\bibfnamefont {J.}~\bibnamefont
  {Vermant}},\ }\bibfield  {title} {\bibinfo {title} {{Direct visualization of
  yielding in model two-dimensional colloidal gels subjected to shear flow}},\
  }\href {https://doi.org/10.1122/1.3237154} {\bibfield  {journal} {\bibinfo
  {journal} {Journal of Rheology}\ }\textbf {\bibinfo {volume} {53}},\ \bibinfo
  {pages} {1437} (\bibinfo {year} {2009})}\BibitemShut {NoStop}%
\bibitem [{\citenamefont {Hoekstra}\ \emph {et~al.}(2005)\citenamefont
  {Hoekstra}, \citenamefont {Mewis}, \citenamefont {Narayanan},\ and\
  \citenamefont {Vermant}}]{Hoekstra:2005}%
  \BibitemOpen
  \bibfield  {author} {\bibinfo {author} {\bibfnamefont {H.}~\bibnamefont
  {Hoekstra}}, \bibinfo {author} {\bibfnamefont {J.}~\bibnamefont {Mewis}},
  \bibinfo {author} {\bibfnamefont {T.}~\bibnamefont {Narayanan}},\ and\
  \bibinfo {author} {\bibfnamefont {J.}~\bibnamefont {Vermant}},\ }\bibfield
  {title} {\bibinfo {title} {Multi length scale analysis of the microstructure
  in sticky sphere dispersions during shear flow},\ }\href@noop {} {\bibfield
  {journal} {\bibinfo  {journal} {Langmuir}\ }\textbf {\bibinfo {volume}
  {21}},\ \bibinfo {pages} {11017} (\bibinfo {year} {2005})}\BibitemShut
  {NoStop}%
\bibitem [{\citenamefont {Zaccone}\ \emph {et~al.}(2009)\citenamefont
  {Zaccone}, \citenamefont {Soos}, \citenamefont {Lattuada}, \citenamefont
  {Wu}, \citenamefont {B{\"A}BLER},\ and\ \citenamefont
  {Morbidelli}}]{Zaccone2009a}%
  \BibitemOpen
  \bibfield  {author} {\bibinfo {author} {\bibfnamefont {A.}~\bibnamefont
  {Zaccone}}, \bibinfo {author} {\bibfnamefont {M.}~\bibnamefont {Soos}},
  \bibinfo {author} {\bibfnamefont {M.}~\bibnamefont {Lattuada}}, \bibinfo
  {author} {\bibfnamefont {H.}~\bibnamefont {Wu}}, \bibinfo {author}
  {\bibfnamefont {M.~U.}\ \bibnamefont {B{\"A}BLER}},\ and\ \bibinfo {author}
  {\bibfnamefont {M.}~\bibnamefont {Morbidelli}},\ }\bibfield  {title}
  {\bibinfo {title} {Breakup of dense colloidal aggregates under hydrodynamic
  stresses},\ }\href@noop {} {\bibfield  {journal} {\bibinfo  {journal}
  {Physical Review E}\ }\textbf {\bibinfo {volume} {79}},\ \bibinfo {pages}
  {061401} (\bibinfo {year} {2009})}\BibitemShut {NoStop}%
\bibitem [{\citenamefont {Torres}\ \emph
  {et~al.}(1991{\natexlab{a}})\citenamefont {Torres}, \citenamefont {Russel},\
  and\ \citenamefont {Schowalter}}]{Torres1991a}%
  \BibitemOpen
  \bibfield  {author} {\bibinfo {author} {\bibfnamefont {F.~E.}\ \bibnamefont
  {Torres}}, \bibinfo {author} {\bibfnamefont {W.~B.}\ \bibnamefont {Russel}},\
  and\ \bibinfo {author} {\bibfnamefont {W.~R.}\ \bibnamefont {Schowalter}},\
  }\bibfield  {title} {\bibinfo {title} {{Floc Structure and Growth Kinetics
  for Rapid Shear Coagulationof Polystyrene Colloids}},\ }\href
  {https://doi.org/10.1016/0021-9797(91)90086-N} {\bibfield  {journal}
  {\bibinfo  {journal} {Journal of Colloid and Interface Science}\ }\textbf
  {\bibinfo {volume} {142}},\ \bibinfo {pages} {554} (\bibinfo {year}
  {1991}{\natexlab{a}})}\BibitemShut {NoStop}%
\bibitem [{\citenamefont {Hunter}\ and\ \citenamefont
  {Frayne}(1980)}]{Hunter1980}%
  \BibitemOpen
  \bibfield  {author} {\bibinfo {author} {\bibfnamefont {R.~J.}\ \bibnamefont
  {Hunter}}\ and\ \bibinfo {author} {\bibfnamefont {J.}~\bibnamefont
  {Frayne}},\ }\bibfield  {title} {\bibinfo {title} {{Flow Behavior of
  Coagulated Colloidal Sols}},\ }\href
  {https://doi.org/10.1016/0021-9797(80)90275-1} {\bibfield  {journal}
  {\bibinfo  {journal} {Journal of Colloid and Interface Science}\ }\textbf
  {\bibinfo {volume} {76}},\ \bibinfo {pages} {107} (\bibinfo {year}
  {1980})}\BibitemShut {NoStop}%
\bibitem [{\citenamefont {Higashitani}\ and\ \citenamefont
  {Iimura}(1998)}]{Higashitani1998}%
  \BibitemOpen
  \bibfield  {author} {\bibinfo {author} {\bibfnamefont {K.}~\bibnamefont
  {Higashitani}}\ and\ \bibinfo {author} {\bibfnamefont {K.}~\bibnamefont
  {Iimura}},\ }\bibfield  {title} {\bibinfo {title} {{Two-Dimensional
  Simulation of the Breakup Process of Aggregates in Shear and Elongational
  Flows}},\ }\href {https://doi.org/10.1006/jcis.1998.5561} {\bibfield
  {journal} {\bibinfo  {journal} {Journal of Colloid and Interface Science}\
  }\textbf {\bibinfo {volume} {204}},\ \bibinfo {pages} {320} (\bibinfo {year}
  {1998})}\BibitemShut {NoStop}%
\bibitem [{\citenamefont {Higashitani}\ \emph {et~al.}(2001)\citenamefont
  {Higashitani}, \citenamefont {Iimura},\ and\ \citenamefont
  {Sanda}}]{Higashitani2001}%
  \BibitemOpen
  \bibfield  {author} {\bibinfo {author} {\bibfnamefont {K.}~\bibnamefont
  {Higashitani}}, \bibinfo {author} {\bibfnamefont {K.}~\bibnamefont
  {Iimura}},\ and\ \bibinfo {author} {\bibfnamefont {H.}~\bibnamefont
  {Sanda}},\ }\bibfield  {title} {\bibinfo {title} {{Simulation of deformation
  and breakupof large aggregates in Flows of viscous Fluids}},\ }\href
  {https://doi.org/10.1016/S0009-2509(00)00477-2} {\bibfield  {journal}
  {\bibinfo  {journal} {Chemical Engineering Science}\ }\textbf {\bibinfo
  {volume} {56}},\ \bibinfo {pages} {2927} (\bibinfo {year}
  {2001})}\BibitemShut {NoStop}%
\bibitem [{\citenamefont {Massaro}\ \emph {et~al.}(2020)\citenamefont
  {Massaro}, \citenamefont {Colombo}, \citenamefont {Van~Puyvelde},\ and\
  \citenamefont {Vermant}}]{Massaro2020}%
  \BibitemOpen
  \bibfield  {author} {\bibinfo {author} {\bibfnamefont {R.}~\bibnamefont
  {Massaro}}, \bibinfo {author} {\bibfnamefont {G.}~\bibnamefont {Colombo}},
  \bibinfo {author} {\bibfnamefont {P.}~\bibnamefont {Van~Puyvelde}},\ and\
  \bibinfo {author} {\bibfnamefont {J.}~\bibnamefont {Vermant}},\ }\bibfield
  {title} {\bibinfo {title} {Viscoelastic cluster densification in sheared
  colloidal gels},\ }\href {https://doi.org/10.1039/C9SM02368B} {\bibfield
  {journal} {\bibinfo  {journal} {Soft Matter}\ }\textbf {\bibinfo {volume}
  {16}},\ \bibinfo {pages} {2437} (\bibinfo {year} {2020})}\BibitemShut
  {NoStop}%
\bibitem [{\citenamefont {Brady}\ and\ \citenamefont
  {Bossis}(1985)}]{Brady:1985}%
  \BibitemOpen
  \bibfield  {author} {\bibinfo {author} {\bibfnamefont {J.~F.}\ \bibnamefont
  {Brady}}\ and\ \bibinfo {author} {\bibfnamefont {G.}~\bibnamefont {Bossis}},\
  }\bibfield  {title} {\bibinfo {title} {The rheology of concentrated
  suspensions of spheres in simple shear flow by numerical simulation},\
  }\href@noop {} {\bibfield  {journal} {\bibinfo  {journal} {Journal of Fluid
  Mechanics}\ }\textbf {\bibinfo {volume} {155}},\ \bibinfo {pages} {105}
  (\bibinfo {year} {1985})}\BibitemShut {NoStop}%
\bibitem [{\citenamefont {Morris}(2009)}]{Morris:2009}%
  \BibitemOpen
  \bibfield  {author} {\bibinfo {author} {\bibfnamefont {J.~F.}\ \bibnamefont
  {Morris}},\ }\bibfield  {title} {\bibinfo {title} {A review of microstructure
  in concentrated suspensions and its implications for rheology and bulk
  flow},\ }\href@noop {} {\bibfield  {journal} {\bibinfo  {journal} {Rheologica
  Acta}\ }\textbf {\bibinfo {volume} {48}},\ \bibinfo {pages} {909} (\bibinfo
  {year} {2009})}\BibitemShut {NoStop}%
\bibitem [{\citenamefont {Varga}\ \emph {et~al.}(2015)\citenamefont {Varga},
  \citenamefont {Wang},\ and\ \citenamefont {Swan}}]{Varga2015a}%
  \BibitemOpen
  \bibfield  {author} {\bibinfo {author} {\bibfnamefont {Z.}~\bibnamefont
  {Varga}}, \bibinfo {author} {\bibfnamefont {G.}~\bibnamefont {Wang}},\ and\
  \bibinfo {author} {\bibfnamefont {J.}~\bibnamefont {Swan}},\ }\bibfield
  {title} {\bibinfo {title} {The hydrodynamics of colloidal gelation},\ }\href
  {https://doi.org/10.1039/C5SM01414J} {\bibfield  {journal} {\bibinfo
  {journal} {Soft Matter}\ }\textbf {\bibinfo {volume} {11}},\ \bibinfo {pages}
  {9009} (\bibinfo {year} {2015})}\BibitemShut {NoStop}%
\bibitem [{\citenamefont {Varga}\ and\ \citenamefont
  {Swan}(2015)}]{Varga2015b}%
  \BibitemOpen
  \bibfield  {author} {\bibinfo {author} {\bibfnamefont {Z.}~\bibnamefont
  {Varga}}\ and\ \bibinfo {author} {\bibfnamefont {J.~W.}\ \bibnamefont
  {Swan}},\ }\bibfield  {title} {\bibinfo {title} {Linear viscoelasticity of
  attractive colloidal dispersions},\ }\href
  {https://doi.org/10.1122/1.4928951} {\bibfield  {journal} {\bibinfo
  {journal} {Journal of Rheology}\ }\textbf {\bibinfo {volume} {59}},\ \bibinfo
  {pages} {1271} (\bibinfo {year} {2015})},\ \Eprint
  {https://arxiv.org/abs/https://doi.org/10.1122/1.4928951}
  {https://doi.org/10.1122/1.4928951} \BibitemShut {NoStop}%
\bibitem [{\citenamefont {Varga}\ and\ \citenamefont {Swan}(2018)}]{Varga2018}%
  \BibitemOpen
  \bibfield  {author} {\bibinfo {author} {\bibfnamefont {Z.}~\bibnamefont
  {Varga}}\ and\ \bibinfo {author} {\bibfnamefont {J.~W.}\ \bibnamefont
  {Swan}},\ }\bibfield  {title} {\bibinfo {title} {Large scale anisotropies in
  sheared colloidal gels},\ }\href {https://doi.org/10.1122/1.5003364}
  {\bibfield  {journal} {\bibinfo  {journal} {Journal of Rheology}\ }\textbf
  {\bibinfo {volume} {62}},\ \bibinfo {pages} {405} (\bibinfo {year} {2018})},\
  \Eprint {https://arxiv.org/abs/https://doi.org/10.1122/1.5003364}
  {https://doi.org/10.1122/1.5003364} \BibitemShut {NoStop}%
\bibitem [{\citenamefont {Varga}\ \emph {et~al.}(2019)\citenamefont {Varga},
  \citenamefont {Grenard}, \citenamefont {Pecorario}, \citenamefont {Taberlet},
  \citenamefont {Dolique}, \citenamefont {Manneville}, \citenamefont {Divoux},
  \citenamefont {McKinley},\ and\ \citenamefont {Swan}}]{Varga2019}%
  \BibitemOpen
  \bibfield  {author} {\bibinfo {author} {\bibfnamefont {Z.}~\bibnamefont
  {Varga}}, \bibinfo {author} {\bibfnamefont {V.}~\bibnamefont {Grenard}},
  \bibinfo {author} {\bibfnamefont {S.}~\bibnamefont {Pecorario}}, \bibinfo
  {author} {\bibfnamefont {N.}~\bibnamefont {Taberlet}}, \bibinfo {author}
  {\bibfnamefont {V.}~\bibnamefont {Dolique}}, \bibinfo {author} {\bibfnamefont
  {S.}~\bibnamefont {Manneville}}, \bibinfo {author} {\bibfnamefont
  {T.}~\bibnamefont {Divoux}}, \bibinfo {author} {\bibfnamefont {G.~H.}\
  \bibnamefont {McKinley}},\ and\ \bibinfo {author} {\bibfnamefont {J.~W.}\
  \bibnamefont {Swan}},\ }\bibfield  {title} {\bibinfo {title} {{Hydrodynamics
  control shear-induced pattern formation in attractive suspensions}},\ }\href
  {https://doi.org/10.1073/pnas.1901370116} {\bibfield  {journal} {\bibinfo
  {journal} {Proceedings of the National Academy of Sciences}\ }\textbf
  {\bibinfo {volume} {116}},\ \bibinfo {pages} {12193} (\bibinfo {year}
  {2019})}\BibitemShut {NoStop}%
\bibitem [{\citenamefont {Lorenzo}\ and\ \citenamefont
  {Marco}(2022)}]{Lorenzo2022}%
  \BibitemOpen
  \bibfield  {author} {\bibinfo {author} {\bibfnamefont {T.}~\bibnamefont
  {Lorenzo}}\ and\ \bibinfo {author} {\bibfnamefont {L.}~\bibnamefont
  {Marco}},\ }\bibfield  {title} {\bibinfo {title} {{Brownian Dynamics
  simulations of shear-induced aggregation of charged colloidal particles in
  the presence of hydrodynamic interactions}},\ }\bibfield  {journal} {\bibinfo
   {journal} {Journal of Colloid and Interface Science}\ }\href
  {https://doi.org/10.1016/j.jcis.2022.05.047} {10.1016/j.jcis.2022.05.047}
  (\bibinfo {year} {2022})\BibitemShut {NoStop}%
\bibitem [{\citenamefont {Jungblut}\ \emph {et~al.}(2019)\citenamefont
  {Jungblut}, \citenamefont {Joswig},\ and\ \citenamefont
  {Eychm\"uller}}]{Jungblut2019}%
  \BibitemOpen
  \bibfield  {author} {\bibinfo {author} {\bibfnamefont {S.}~\bibnamefont
  {Jungblut}}, \bibinfo {author} {\bibfnamefont {J.-O.}\ \bibnamefont
  {Joswig}},\ and\ \bibinfo {author} {\bibfnamefont {A.}~\bibnamefont
  {Eychm\"uller}},\ }\bibfield  {title} {\bibinfo {title} {Diffusion- and
  reaction-limited cluster aggregation revisited},\ }\href
  {https://doi.org/10.1039/C9CP00549H} {\bibfield  {journal} {\bibinfo
  {journal} {Physical Chemistry Chemical Physics}\ }\textbf {\bibinfo {volume}
  {21}},\ \bibinfo {pages} {5723} (\bibinfo {year} {2019})}\BibitemShut
  {NoStop}%
\bibitem [{\citenamefont {Colombo}\ and\ \citenamefont
  {Del~Gado}(2014{\natexlab{a}})}]{Colombo2014a}%
  \BibitemOpen
  \bibfield  {author} {\bibinfo {author} {\bibfnamefont {J.}~\bibnamefont
  {Colombo}}\ and\ \bibinfo {author} {\bibfnamefont {E.}~\bibnamefont
  {Del~Gado}},\ }\bibfield  {title} {\bibinfo {title} {Stress localization,
  stiffening, and yielding in a model colloidal gel},\ }\href
  {https://doi.org/10.1122/1.4882021} {\bibfield  {journal} {\bibinfo
  {journal} {Journal of Rheology}\ }\textbf {\bibinfo {volume} {58}},\ \bibinfo
  {pages} {1089} (\bibinfo {year} {2014}{\natexlab{a}})}\BibitemShut {NoStop}%
\bibitem [{\citenamefont {Bouzid}\ \emph {et~al.}(2018)\citenamefont {Bouzid},
  \citenamefont {Keshavarz}, \citenamefont {Geri}, \citenamefont {Divoux},
  \citenamefont {Del~Gado},\ and\ \citenamefont {McKinley}}]{Bouzid2018a}%
  \BibitemOpen
  \bibfield  {author} {\bibinfo {author} {\bibfnamefont {M.}~\bibnamefont
  {Bouzid}}, \bibinfo {author} {\bibfnamefont {B.}~\bibnamefont {Keshavarz}},
  \bibinfo {author} {\bibfnamefont {M.}~\bibnamefont {Geri}}, \bibinfo {author}
  {\bibfnamefont {T.}~\bibnamefont {Divoux}}, \bibinfo {author} {\bibfnamefont
  {E.}~\bibnamefont {Del~Gado}},\ and\ \bibinfo {author} {\bibfnamefont
  {G.~H.}\ \bibnamefont {McKinley}},\ }\bibfield  {title} {\bibinfo {title}
  {{Computing the linear viscoelastic properties of soft gels using an
  optimally windowed chirp protocol}},\ }\href
  {https://doi.org/10.1122/1.5018715} {\bibfield  {journal} {\bibinfo
  {journal} {Journal of Rheology}\ }\textbf {\bibinfo {volume} {62}},\ \bibinfo
  {pages} {1037} (\bibinfo {year} {2018})}\BibitemShut {NoStop}%
\bibitem [{\citenamefont {Kimbonguila~Manounou}\ and\ \citenamefont
  {R\'emond}(2014)}]{Kimbonguila2014}%
  \BibitemOpen
  \bibfield  {author} {\bibinfo {author} {\bibfnamefont {A.}~\bibnamefont
  {Kimbonguila~Manounou}}\ and\ \bibinfo {author} {\bibfnamefont
  {S.}~\bibnamefont {R\'emond}},\ }\bibfield  {title} {\bibinfo {title}
  {{Discrete element modeling of the microstructure of fine particle
  agglomerates in sheared dilute suspension}},\ }\href
  {https://doi.org/10.1016/j.physa.2014.06.023} {\bibfield  {journal} {\bibinfo
   {journal} {Physica A}\ }\textbf {\bibinfo {volume} {412}},\ \bibinfo {pages}
  {66} (\bibinfo {year} {2014})}\BibitemShut {NoStop}%
\bibitem [{\citenamefont {Eggersdorfer}\ \emph {et~al.}(2010)\citenamefont
  {Eggersdorfer}, \citenamefont {Kadau}, \citenamefont {Herrmann},\ and\
  \citenamefont {Pratsinis}}]{Eggersdorfer2010}%
  \BibitemOpen
  \bibfield  {author} {\bibinfo {author} {\bibfnamefont {M.~L.}\ \bibnamefont
  {Eggersdorfer}}, \bibinfo {author} {\bibfnamefont {D.}~\bibnamefont {Kadau}},
  \bibinfo {author} {\bibfnamefont {H.~J.}\ \bibnamefont {Herrmann}},\ and\
  \bibinfo {author} {\bibfnamefont {S.~E.}\ \bibnamefont {Pratsinis}},\
  }\bibfield  {title} {\bibinfo {title} {{Fragmentation and restructuring of
  soft-agglomerates under shear}},\ }\href
  {https://doi.org/10.1016/j.jcis.2009.10.062} {\bibfield  {journal} {\bibinfo
  {journal} {Journal of Colloid and Interface Science}\ }\textbf {\bibinfo
  {volume} {342}},\ \bibinfo {pages} {261} (\bibinfo {year}
  {2010})}\BibitemShut {NoStop}%
\bibitem [{\citenamefont {Ruan}\ \emph {et~al.}(2020)\citenamefont {Ruan},
  \citenamefont {Chen},\ and\ \citenamefont {Li}}]{Ruan2020}%
  \BibitemOpen
  \bibfield  {author} {\bibinfo {author} {\bibfnamefont {X.}~\bibnamefont
  {Ruan}}, \bibinfo {author} {\bibfnamefont {S.}~\bibnamefont {Chen}},\ and\
  \bibinfo {author} {\bibfnamefont {S.}~\bibnamefont {Li}},\ }\bibfield
  {title} {\bibinfo {title} {{Structural evolution and breakage of dense
  agglomerates in shear flowand Taylor-Green vortex}},\ }\href
  {https://doi.org/10.1016/j.ces.2019.115261} {\bibfield  {journal} {\bibinfo
  {journal} {Chemical Engineering Science}\ }\textbf {\bibinfo {volume}
  {211}},\ \bibinfo {pages} {115261} (\bibinfo {year} {2020})}\BibitemShut
  {NoStop}%
\bibitem [{\citenamefont {Soos}\ \emph {et~al.}(2006)\citenamefont {Soos},
  \citenamefont {Sefcik},\ and\ \citenamefont {Morbidelli}}]{Soos2006}%
  \BibitemOpen
  \bibfield  {author} {\bibinfo {author} {\bibfnamefont {M.}~\bibnamefont
  {Soos}}, \bibinfo {author} {\bibfnamefont {J.}~\bibnamefont {Sefcik}},\ and\
  \bibinfo {author} {\bibfnamefont {M.}~\bibnamefont {Morbidelli}},\ }\bibfield
   {title} {\bibinfo {title} {Investigation of aggregation, breakage and
  restructuring kinetics of colloidal dispersions in turbulent flows by
  population balance modeling and static light scattering},\ }\href@noop {}
  {\bibfield  {journal} {\bibinfo  {journal} {Chemical Engineering Science}\
  }\textbf {\bibinfo {volume} {61}},\ \bibinfo {pages} {2349} (\bibinfo {year}
  {2006})}\BibitemShut {NoStop}%
\bibitem [{\citenamefont {Soos}\ \emph {et~al.}(2007)\citenamefont {Soos},
  \citenamefont {Wang}, \citenamefont {Fox}, \citenamefont {Sefcik},\ and\
  \citenamefont {Morbidelli}}]{Soos2007}%
  \BibitemOpen
  \bibfield  {author} {\bibinfo {author} {\bibfnamefont {M.}~\bibnamefont
  {Soos}}, \bibinfo {author} {\bibfnamefont {L.}~\bibnamefont {Wang}}, \bibinfo
  {author} {\bibfnamefont {R.~O.}\ \bibnamefont {Fox}}, \bibinfo {author}
  {\bibfnamefont {J.}~\bibnamefont {Sefcik}},\ and\ \bibinfo {author}
  {\bibfnamefont {M.}~\bibnamefont {Morbidelli}},\ }\bibfield  {title}
  {\bibinfo {title} {Population balance modeling of aggregation and breakage in
  turbulent taylor--couette flow},\ }\href@noop {} {\bibfield  {journal}
  {\bibinfo  {journal} {Journal of Colloid and Interface Science}\ }\textbf
  {\bibinfo {volume} {307}},\ \bibinfo {pages} {433} (\bibinfo {year}
  {2007})}\BibitemShut {NoStop}%
\bibitem [{\citenamefont {Puisto}\ \emph {et~al.}(2012)\citenamefont {Puisto},
  \citenamefont {Illa}, \citenamefont {Mohtaschemi},\ and\ \citenamefont
  {Alava}}]{Puisto2012}%
  \BibitemOpen
  \bibfield  {author} {\bibinfo {author} {\bibfnamefont {A.}~\bibnamefont
  {Puisto}}, \bibinfo {author} {\bibfnamefont {X.}~\bibnamefont {Illa}},
  \bibinfo {author} {\bibfnamefont {M.}~\bibnamefont {Mohtaschemi}},\ and\
  \bibinfo {author} {\bibfnamefont {M.~J.}\ \bibnamefont {Alava}},\ }\bibfield
  {title} {\bibinfo {title} {Modeling the viscosity and aggregation of
  suspensions of highly anisotropic nanoparticles},\ }\href
  {https://doi.org/10.1140/epje/i2012-12006-1} {\bibfield  {journal} {\bibinfo
  {journal} {The European Physical Journal E}\ }\textbf {\bibinfo {volume}
  {35}},\ \bibinfo {pages} {1} (\bibinfo {year} {2012})}\BibitemShut {NoStop}%
\bibitem [{\citenamefont {Lattuada}\ \emph {et~al.}(2016)\citenamefont
  {Lattuada}, \citenamefont {Zaccone}, \citenamefont {Wu},\ and\ \citenamefont
  {Morbidelli}}]{Lattuada2016}%
  \BibitemOpen
  \bibfield  {author} {\bibinfo {author} {\bibfnamefont {M.}~\bibnamefont
  {Lattuada}}, \bibinfo {author} {\bibfnamefont {A.}~\bibnamefont {Zaccone}},
  \bibinfo {author} {\bibfnamefont {H.}~\bibnamefont {Wu}},\ and\ \bibinfo
  {author} {\bibfnamefont {M.}~\bibnamefont {Morbidelli}},\ }\bibfield  {title}
  {\bibinfo {title} {{Population-balance description of shear-induced
  clustering, gelation and suspension viscosity in sheared DLVO colloids}},\
  }\href {https://doi.org/10.1039/C6SM01097K} {\bibfield  {journal} {\bibinfo
  {journal} {Soft Matter}\ }\textbf {\bibinfo {volume} {12}},\ \bibinfo {pages}
  {5313} (\bibinfo {year} {2016})}\BibitemShut {NoStop}%
\bibitem [{\citenamefont {Banasiak}\ \emph
  {et~al.}(2020{\natexlab{a}})\citenamefont {Banasiak}, \citenamefont {Lamb},\
  and\ \citenamefont {Lauren\c{c}ot}}]{Banasiak2020a}%
  \BibitemOpen
  \bibfield  {author} {\bibinfo {author} {\bibfnamefont {J.}~\bibnamefont
  {Banasiak}}, \bibinfo {author} {\bibfnamefont {W.}~\bibnamefont {Lamb}},\
  and\ \bibinfo {author} {\bibfnamefont {P.}~\bibnamefont {Lauren\c{c}ot}},\
  }\href@noop {} {\emph {\bibinfo {title} {{Analytic Methods for
  Coagulation-Fragmentation Models, Volume I}}}},\ edited by\ \bibinfo {editor}
  {\bibfnamefont {J.}~\bibnamefont {Banasiak}}, \bibinfo {editor}
  {\bibfnamefont {W.}~\bibnamefont {Lamb}},\ and\ \bibinfo {editor}
  {\bibfnamefont {P.}~\bibnamefont {Lauren\c{c}ot}},\ \bibinfo {series}
  {Monographs and Research Notes in Mathematics}, Vol.~\bibinfo {volume} {1}\
  (\bibinfo  {publisher} {CRC Press},\ \bibinfo {address} {CRC PressTaylor \&
  Francis Group6000 Broken Sound Parkway NW, Suite 300Boca Raton, FL
  33487-2742},\ \bibinfo {year} {2020})\BibitemShut {NoStop}%
\bibitem [{\citenamefont {Kantor}\ and\ \citenamefont
  {Witten}(1984)}]{Kantor1984a}%
  \BibitemOpen
  \bibfield  {author} {\bibinfo {author} {\bibfnamefont {Y.}~\bibnamefont
  {Kantor}}\ and\ \bibinfo {author} {\bibfnamefont {T.~A.}\ \bibnamefont
  {Witten}},\ }\bibfield  {title} {\bibinfo {title} {Mechanical stability of
  tenuous objects},\ }\href
  {https://doi.org/10.1051/jphyslet:019840045013067500} {\bibfield  {journal}
  {\bibinfo  {journal} {Journal de Physique Lettres}\ }\textbf {\bibinfo
  {volume} {45}},\ \bibinfo {pages} {675} (\bibinfo {year} {1984})}\BibitemShut
  {NoStop}%
\bibitem [{\citenamefont {Mellema}\ \emph {et~al.}(2002)\citenamefont
  {Mellema}, \citenamefont {van Opheusde},\ and\ \citenamefont {van
  Vliet}}]{Mellema2002}%
  \BibitemOpen
  \bibfield  {author} {\bibinfo {author} {\bibfnamefont {M.}~\bibnamefont
  {Mellema}}, \bibinfo {author} {\bibfnamefont {J.~H.~J.}\ \bibnamefont {van
  Opheusde}},\ and\ \bibinfo {author} {\bibfnamefont {T.}~\bibnamefont {van
  Vliet}},\ }\bibfield  {title} {\bibinfo {title} {Categorization of
  rheological scaling models for particlegels applied to casein gels},\ }\href
  {https://doi.org/10.1122/1.1423311} {\bibfield  {journal} {\bibinfo
  {journal} {Journal of Rheology}\ }\textbf {\bibinfo {volume} {46}},\ \bibinfo
  {pages} {11} (\bibinfo {year} {2002})}\BibitemShut {NoStop}%
\bibitem [{\citenamefont {Brakalov}(1987)}]{Brakalov1987}%
  \BibitemOpen
  \bibfield  {author} {\bibinfo {author} {\bibfnamefont {L.~B.}\ \bibnamefont
  {Brakalov}},\ }\bibfield  {title} {\bibinfo {title} {Connection between the
  orthokinetic coagulation capture efficiency of aggregates and their maximum
  size},\ }\href {https://doi.org/10.1016/0009-2509(87)80111-2} {\bibfield
  {journal} {\bibinfo  {journal} {Chemical Engineering Science}\ }\textbf
  {\bibinfo {volume} {42}},\ \bibinfo {pages} {2373} (\bibinfo {year}
  {1987})}\BibitemShut {NoStop}%
\bibitem [{\citenamefont {Torres}\ \emph
  {et~al.}(1991{\natexlab{b}})\citenamefont {Torres}, \citenamefont {Russel},\
  and\ \citenamefont {Schowalter}}]{Torres1991b}%
  \BibitemOpen
  \bibfield  {author} {\bibinfo {author} {\bibfnamefont {F.~E.}\ \bibnamefont
  {Torres}}, \bibinfo {author} {\bibfnamefont {W.~B.}\ \bibnamefont {Russel}},\
  and\ \bibinfo {author} {\bibfnamefont {W.~R.}\ \bibnamefont {Schowalter}},\
  }\bibfield  {title} {\bibinfo {title} {{Simulation of Coagulation in Viscous
  Flows}},\ }\href {https://doi.org/10.1016/0021-9797(91)90099-T} {\bibfield
  {journal} {\bibinfo  {journal} {Journal of Colloid and Interface Science}\
  }\textbf {\bibinfo {volume} {145}},\ \bibinfo {pages} {51} (\bibinfo {year}
  {1991}{\natexlab{b}})}\BibitemShut {NoStop}%
\bibitem [{\citenamefont {Snabre}\ and\ \citenamefont
  {Mills}(1996)}]{Snabre1996}%
  \BibitemOpen
  \bibfield  {author} {\bibinfo {author} {\bibfnamefont {P.}~\bibnamefont
  {Snabre}}\ and\ \bibinfo {author} {\bibfnamefont {P.}~\bibnamefont {Mills}},\
  }\bibfield  {title} {\bibinfo {title} {{I. Rheology of Weakly Flocculated
  Suspensions of Rigid Particles}},\ }\href
  {https://doi.org/10.1051/jp3:1996215} {\bibfield  {journal} {\bibinfo
  {journal} {Journal de Physique III}\ }\textbf {\bibinfo {volume} {6}},\
  \bibinfo {pages} {1811} (\bibinfo {year} {1996})}\BibitemShut {NoStop}%
\bibitem [{\citenamefont {Marshall}\ and\ \citenamefont
  {Li}(2014)}]{Marshall2014}%
  \BibitemOpen
  \bibfield  {author} {\bibinfo {author} {\bibfnamefont {J.~S.}\ \bibnamefont
  {Marshall}}\ and\ \bibinfo {author} {\bibfnamefont {S.}~\bibnamefont {Li}},\
  }\href {https://doi.org/10.1017/CBO9781139424547} {\emph {\bibinfo {title}
  {{Adhesive Particle Flow: A Discrete-Element Approach}}}},\ \bibinfo
  {edition} {1st}\ ed.,\ edited by\ \bibinfo {editor} {\bibfnamefont {J.~S.}\
  \bibnamefont {Marshall}}\ and\ \bibinfo {editor} {\bibfnamefont
  {S.}~\bibnamefont {Li}}\ (\bibinfo  {publisher} {Cambridge University
  Press},\ \bibinfo {address} {32 Avenue of the Americas, NY 10013-2473, USA},\
  \bibinfo {year} {2014})\BibitemShut {NoStop}%
\bibitem [{\citenamefont {von Mises}(1913)}]{vonMises1913}%
  \BibitemOpen
  \bibfield  {author} {\bibinfo {author} {\bibfnamefont {R.}~\bibnamefont {von
  Mises}},\ }\bibfield  {title} {\bibinfo {title} {{Mechanik der festen
  K\"orper im plastisch-deformablen Zustand}},\ }\href
  {http://www.digizeitschriften.de/dms/img/?PID=GDZPPN002503697} {\bibfield
  {journal} {\bibinfo  {journal} {Nachrichten von der Gesellschaft der
  Wissenschaften zu G\"ottingen}\ }\textbf {\bibinfo {volume} {1}},\ \bibinfo
  {pages} {582} (\bibinfo {year} {1913})}\BibitemShut {NoStop}%
\bibitem [{\citenamefont {Griffith}(1921)}]{Griffith1921}%
  \BibitemOpen
  \bibfield  {author} {\bibinfo {author} {\bibfnamefont {A.~A.}\ \bibnamefont
  {Griffith}},\ }\bibfield  {title} {\bibinfo {title} {{VI. The phenomena of
  rupture and flow in solids}},\ }\href
  {https://doi.org/10.1098/rsta.1921.0006} {\bibfield  {journal} {\bibinfo
  {journal} {Philosophical Transactions of the Royal Society A}\ }\textbf
  {\bibinfo {volume} {221}},\ \bibinfo {pages} {582} (\bibinfo {year}
  {1921})}\BibitemShut {NoStop}%
\bibitem [{\citenamefont {Irwin}(1957)}]{Irwin1957}%
  \BibitemOpen
  \bibfield  {author} {\bibinfo {author} {\bibfnamefont {G.~R.}\ \bibnamefont
  {Irwin}},\ }\bibfield  {title} {\bibinfo {title} {{Analysis of Stresses and
  Strains Near the End of a Crack Traversing a Plate}},\ }\href
  {https://doi.org/10.1115/1.4011547} {\bibfield  {journal} {\bibinfo
  {journal} {Journal of Applied Mechanics}\ }\textbf {\bibinfo {volume} {24}},\
  \bibinfo {pages} {361} (\bibinfo {year} {1957})}\BibitemShut {NoStop}%
\bibitem [{\citenamefont {Creton}\ and\ \citenamefont
  {Ciccotti}(2016)}]{Creton2016}%
  \BibitemOpen
  \bibfield  {author} {\bibinfo {author} {\bibfnamefont {C.}~\bibnamefont
  {Creton}}\ and\ \bibinfo {author} {\bibfnamefont {M.}~\bibnamefont
  {Ciccotti}},\ }\bibfield  {title} {\bibinfo {title} {Fracture and adhesion of
  soft materials: a review},\ }\href
  {https://doi.org/10.1088/0034-4885/79/4/046601} {\bibfield  {journal}
  {\bibinfo  {journal} {Reports on Progress in Physics}\ }\textbf {\bibinfo
  {volume} {79}},\ \bibinfo {pages} {046601} (\bibinfo {year}
  {2016})}\BibitemShut {NoStop}%
\bibitem [{\citenamefont {Buckingham}(1914)}]{Buckingham1914}%
  \BibitemOpen
  \bibfield  {author} {\bibinfo {author} {\bibfnamefont {E.}~\bibnamefont
  {Buckingham}},\ }\bibfield  {title} {\bibinfo {title} {{On Physically Similar
  Systems; Illustrations of the Use of Dimensional Equations}},\ }\href
  {https://doi.org/10.1103/PhysRev.4.345} {\bibfield  {journal} {\bibinfo
  {journal} {Physical Review}\ }\textbf {\bibinfo {volume} {4}},\ \bibinfo
  {pages} {345} (\bibinfo {year} {1914})}\BibitemShut {NoStop}%
\bibitem [{\citenamefont {Buckingham}(1915{\natexlab{a}})}]{Buckingham1915a}%
  \BibitemOpen
  \bibfield  {author} {\bibinfo {author} {\bibfnamefont {E.}~\bibnamefont
  {Buckingham}},\ }\bibfield  {title} {\bibinfo {title} {{The Principle of
  Similitude}},\ }\href {https://doi.org/10.1038/096396d0} {\bibfield
  {journal} {\bibinfo  {journal} {Nature}\ }\textbf {\bibinfo {volume} {96}},\
  \bibinfo {pages} {396} (\bibinfo {year} {1915}{\natexlab{a}})}\BibitemShut
  {NoStop}%
\bibitem [{\citenamefont {Buckingham}(1915{\natexlab{b}})}]{Buckingham1915b}%
  \BibitemOpen
  \bibfield  {author} {\bibinfo {author} {\bibfnamefont {E.}~\bibnamefont
  {Buckingham}},\ }\bibfield  {title} {\bibinfo {title} {Model experiments and
  the forms of empirical equations},\ }in\ \href@noop {} {\emph {\bibinfo
  {booktitle} {Transactions of the American Society of Mechanical
  Engineers}}},\ Vol.~\bibinfo {volume} {37},\ \bibinfo {organization}
  {American Society of Mechanical Engineers}\ (\bibinfo  {publisher} {American
  Society of Mechanical Engineers},\ \bibinfo {year} {1915})\ pp.\ \bibinfo
  {pages} {263--296}\BibitemShut {NoStop}%
\bibitem [{\citenamefont {Jamali}\ \emph {et~al.}(2020)\citenamefont {Jamali},
  \citenamefont {Armstrong},\ and\ \citenamefont {McKinley}}]{Jamali2020}%
  \BibitemOpen
  \bibfield  {author} {\bibinfo {author} {\bibfnamefont {S.}~\bibnamefont
  {Jamali}}, \bibinfo {author} {\bibfnamefont {R.~C.}\ \bibnamefont
  {Armstrong}},\ and\ \bibinfo {author} {\bibfnamefont {G.~H.}\ \bibnamefont
  {McKinley}},\ }\bibfield  {title} {\bibinfo {title} {Time-rate-transformation
  framework for targeted assembly of short-range attractive colloidal
  suspensions},\ }\href {https://doi.org/10.1016/j.mtadv.2019.100026}
  {\bibfield  {journal} {\bibinfo  {journal} {Materials Today Advances}\
  }\textbf {\bibinfo {volume} {5}},\ \bibinfo {pages} {100026} (\bibinfo {year}
  {2020})}\BibitemShut {NoStop}%
\bibitem [{\citenamefont {Nabizadeh}\ and\ \citenamefont
  {Jamali}(2021)}]{Nabizadeh2021}%
  \BibitemOpen
  \bibfield  {author} {\bibinfo {author} {\bibfnamefont {M.}~\bibnamefont
  {Nabizadeh}}\ and\ \bibinfo {author} {\bibfnamefont {S.}~\bibnamefont
  {Jamali}},\ }\bibfield  {title} {\bibinfo {title} {Life and death of
  colloidal bonds control the rate-dependent rheology of gels},\ }\href
  {https://doi.org/10.1038/s41467-021-24416-x} {\bibfield  {journal} {\bibinfo
  {journal} {Nature Communications}\ }\textbf {\bibinfo {volume} {12}},\
  \bibinfo {pages} {1} (\bibinfo {year} {2021})}\BibitemShut {NoStop}%
\bibitem [{\citenamefont {Thompson}\ \emph {et~al.}(2022)\citenamefont
  {Thompson}, \citenamefont {Aktulga}, \citenamefont {Berger}, \citenamefont
  {Bolintineanu}, \citenamefont {Brown}, \citenamefont {Crozier}, \citenamefont
  {in~'t Veld}, \citenamefont {Kohlmeyer}, \citenamefont {Moore}, \citenamefont
  {Nguyen}, \citenamefont {Shan}, \citenamefont {Stevens}, \citenamefont
  {Tranchida}, \citenamefont {Trott},\ and\ \citenamefont
  {Plimpton}}]{Thompson2022}%
  \BibitemOpen
  \bibfield  {author} {\bibinfo {author} {\bibfnamefont {A.~P.}\ \bibnamefont
  {Thompson}}, \bibinfo {author} {\bibfnamefont {H.~M.}\ \bibnamefont
  {Aktulga}}, \bibinfo {author} {\bibfnamefont {R.}~\bibnamefont {Berger}},
  \bibinfo {author} {\bibfnamefont {D.~S.}\ \bibnamefont {Bolintineanu}},
  \bibinfo {author} {\bibfnamefont {W.~M.}\ \bibnamefont {Brown}}, \bibinfo
  {author} {\bibfnamefont {P.~S.}\ \bibnamefont {Crozier}}, \bibinfo {author}
  {\bibfnamefont {P.~J.}\ \bibnamefont {in~'t Veld}}, \bibinfo {author}
  {\bibfnamefont {A.}~\bibnamefont {Kohlmeyer}}, \bibinfo {author}
  {\bibfnamefont {S.~G.}\ \bibnamefont {Moore}}, \bibinfo {author}
  {\bibfnamefont {T.~D.}\ \bibnamefont {Nguyen}}, \bibinfo {author}
  {\bibfnamefont {R.}~\bibnamefont {Shan}}, \bibinfo {author} {\bibfnamefont
  {M.~J.}\ \bibnamefont {Stevens}}, \bibinfo {author} {\bibfnamefont
  {J.}~\bibnamefont {Tranchida}}, \bibinfo {author} {\bibfnamefont
  {C.}~\bibnamefont {Trott}},\ and\ \bibinfo {author} {\bibfnamefont {S.~J.}\
  \bibnamefont {Plimpton}},\ }\bibfield  {title} {\bibinfo {title} {{LAMMPS} -
  a flexible simulation tool for particle-based materials modeling at the
  atomic, meso, and continuum scales},\ }\href
  {https://doi.org/10.1016/j.cpc.2021.108171} {\bibfield  {journal} {\bibinfo
  {journal} {Computer Physics Communications}\ }\textbf {\bibinfo {volume}
  {271}},\ \bibinfo {pages} {108171} (\bibinfo {year} {2022})}\BibitemShut
  {NoStop}%
\bibitem [{\citenamefont {Stukowski}(2010)}]{Stukowski2010}%
  \BibitemOpen
  \bibfield  {author} {\bibinfo {author} {\bibfnamefont {A.}~\bibnamefont
  {Stukowski}},\ }\bibfield  {title} {\bibinfo {title} {{Visualization and
  analysis of atomistic simulation data with OVITO-the Open Visualization
  Tool}},\ }\href {https://doi.org/10.1088/0965-0393/18/1/015012} {\bibfield
  {journal} {\bibinfo  {journal} {Modelling and Simulation in Materials Science
  and Engineering}\ }\textbf {\bibinfo {volume} {18}},\ \bibinfo {pages}
  {015012} (\bibinfo {year} {2010})}\BibitemShut {NoStop}%
\bibitem [{Note1()}]{Note1}%
  \BibitemOpen
  \bibinfo {note} {Additional simulations were carried out with various values
  of $C$ and the main results remained similar up to a rescaling of the time
  scale.}\BibitemShut {Stop}%
\bibitem [{\citenamefont {Swope}\ \emph {et~al.}(1982)\citenamefont {Swope},
  \citenamefont {Andersen}, \citenamefont {Berens},\ and\ \citenamefont
  {Wilson}}]{Swope1982}%
  \BibitemOpen
  \bibfield  {author} {\bibinfo {author} {\bibfnamefont {W.~C.}\ \bibnamefont
  {Swope}}, \bibinfo {author} {\bibfnamefont {H.~C.}\ \bibnamefont {Andersen}},
  \bibinfo {author} {\bibfnamefont {P.~H.}\ \bibnamefont {Berens}},\ and\
  \bibinfo {author} {\bibfnamefont {K.~R.}\ \bibnamefont {Wilson}},\ }\bibfield
   {title} {\bibinfo {title} {A computer simulation method for the calculation
  of equilibrium constants for the formation of physical clusters of molecules:
  Application to small water clusters},\ }\href
  {https://doi.org/10.1063/1.442716} {\bibfield  {journal} {\bibinfo  {journal}
  {The Journal of Chemical Physics}\ }\textbf {\bibinfo {volume} {76}},\
  \bibinfo {pages} {637} (\bibinfo {year} {1982})},\ \Eprint
  {https://arxiv.org/abs/https://doi.org/10.1063/1.442716}
  {https://doi.org/10.1063/1.442716} \BibitemShut {NoStop}%
\bibitem [{\citenamefont {Pearce}(2005)}]{Pearce2005}%
  \BibitemOpen
  \bibfield  {author} {\bibinfo {author} {\bibfnamefont {D.~J.}\ \bibnamefont
  {Pearce}},\ }\href
  {http://citeseerx.ist.psu.edu/viewdoc/summary?doi=10.1.1.102.1707} {\emph
  {\bibinfo {title} {An improved algorithm for finding the strongly connected
  components of a directed graph}}},\ \bibinfo {type} {techreport}\ (\bibinfo
  {institution} {Victoria University},\ \bibinfo {year} {2005})\BibitemShut
  {NoStop}%
\bibitem [{\citenamefont {Colombo}\ \emph {et~al.}(2013)\citenamefont
  {Colombo}, \citenamefont {Widmer-Cooper},\ and\ \citenamefont
  {Del~Gado}}]{Colombo2013}%
  \BibitemOpen
  \bibfield  {author} {\bibinfo {author} {\bibfnamefont {J.}~\bibnamefont
  {Colombo}}, \bibinfo {author} {\bibfnamefont {A.}~\bibnamefont
  {Widmer-Cooper}},\ and\ \bibinfo {author} {\bibfnamefont {E.}~\bibnamefont
  {Del~Gado}},\ }\bibfield  {title} {\bibinfo {title} {{Microscopic Picture of
  Cooperative Processes in Restructuring Gel Networks}},\ }\href
  {https://doi.org/10.1103/PhysRevLett.110.198301} {\bibfield  {journal}
  {\bibinfo  {journal} {Physical Review Letters}\ }\textbf {\bibinfo {volume}
  {110}},\ \bibinfo {pages} {198301} (\bibinfo {year} {2013})}\BibitemShut
  {NoStop}%
\bibitem [{\citenamefont {Colombo}\ and\ \citenamefont
  {Del~Gado}(2014{\natexlab{b}})}]{Colombo2014b}%
  \BibitemOpen
  \bibfield  {author} {\bibinfo {author} {\bibfnamefont {J.}~\bibnamefont
  {Colombo}}\ and\ \bibinfo {author} {\bibfnamefont {E.}~\bibnamefont
  {Del~Gado}},\ }\bibfield  {title} {\bibinfo {title} {Self-assembly and
  cooperative dynamics of a model colloidal gel network},\ }\href
  {https://doi.org/10.1039/C4SM00219A} {\bibfield  {journal} {\bibinfo
  {journal} {Soft Matter}\ }\textbf {\bibinfo {volume} {10}},\ \bibinfo {pages}
  {4003} (\bibinfo {year} {2014}{\natexlab{b}})}\BibitemShut {NoStop}%
\bibitem [{\citenamefont {Banasiak}\ \emph
  {et~al.}(2020{\natexlab{b}})\citenamefont {Banasiak}, \citenamefont {Lamb},\
  and\ \citenamefont {Lauren\c{c}ot}}]{Banasiak2020b}%
  \BibitemOpen
  \bibfield  {author} {\bibinfo {author} {\bibfnamefont {J.}~\bibnamefont
  {Banasiak}}, \bibinfo {author} {\bibfnamefont {W.}~\bibnamefont {Lamb}},\
  and\ \bibinfo {author} {\bibfnamefont {P.}~\bibnamefont {Lauren\c{c}ot}},\
  }\href@noop {} {\emph {\bibinfo {title} {{Analytic Methods for
  Coagulation-Fragmentation Models, Volume II}}}},\ edited by\ \bibinfo
  {editor} {\bibfnamefont {J.}~\bibnamefont {Banasiak}}, \bibinfo {editor}
  {\bibfnamefont {W.}~\bibnamefont {Lamb}},\ and\ \bibinfo {editor}
  {\bibfnamefont {P.}~\bibnamefont {Lauren\c{c}ot}},\ \bibinfo {series}
  {Monographs and Research Notes in Mathematics}, Vol.~\bibinfo {volume} {2}\
  (\bibinfo  {publisher} {CRC Press},\ \bibinfo {address} {CRC PressTaylor \&
  Francis Group6000 Broken Sound Parkway NW, Suite 300Boca Raton, FL
  33487-2742},\ \bibinfo {year} {2020})\BibitemShut {NoStop}%
\bibitem [{\citenamefont {Golse}(2005)}]{Golse2005}%
  \BibitemOpen
  \bibfield  {author} {\bibinfo {author} {\bibfnamefont {F.}~\bibnamefont
  {Golse}},\ }\bibinfo {title} {{Handbook of Differential Equations:
  Evolutionary Equations}}\ (\bibinfo  {publisher} {Elsevier-North-Holland},\
  \bibinfo {year} {2005})\ Chap.\ \bibinfo {chapter} {The Boltzmann Equation
  and Its Hydrodynamic Limits}, pp.\ \bibinfo {pages} {159--301}\BibitemShut
  {NoStop}%
\bibitem [{\citenamefont {Alexeev}(2004)}]{Alexeev2004}%
  \BibitemOpen
  \bibfield  {author} {\bibinfo {author} {\bibfnamefont {B.~V.}\ \bibnamefont
  {Alexeev}},\ }\href@noop {} {\emph {\bibinfo {title} {{Generalized Boltzmann
  Physical Kinetics}}}},\ edited by\ \bibinfo {editor} {\bibfnamefont {B.~V.}\
  \bibnamefont {Alexeev}}\ (\bibinfo  {publisher} {Elsevier},\ \bibinfo {year}
  {2004})\BibitemShut {NoStop}%
\bibitem [{\citenamefont {Stadnichuk}\ \emph {et~al.}(2015)\citenamefont
  {Stadnichuk}, \citenamefont {Bodrova},\ and\ \citenamefont
  {Brilliantov}}]{Stadnichuk2015}%
  \BibitemOpen
  \bibfield  {author} {\bibinfo {author} {\bibfnamefont {V.}~\bibnamefont
  {Stadnichuk}}, \bibinfo {author} {\bibfnamefont {A.}~\bibnamefont
  {Bodrova}},\ and\ \bibinfo {author} {\bibfnamefont {N.}~\bibnamefont
  {Brilliantov}},\ }\bibfield  {title} {\bibinfo {title} {{Smoluchowski
  aggregation fragmentation equations: Fast numerical method to find
  steady-state solutions}},\ }\href {https://doi.org/10.1142/S0217979215502082}
  {\bibfield  {journal} {\bibinfo  {journal} {International Journal of Modern
  Physics B}\ }\textbf {\bibinfo {volume} {29}},\ \bibinfo {pages} {1}
  (\bibinfo {year} {2015})}\BibitemShut {NoStop}%
\bibitem [{Note2()}]{Note2}%
  \BibitemOpen
  \bibinfo {note} {Taking some usual examples leads to $\delta \approx \protect
  \qopname \relax o{arg}\protect \qopname \relax m{min}_{d\in \protect \mathbb
  {R}_+^*}w\left (d\right )$}\BibitemShut {NoStop}%
\bibitem [{\citenamefont {Sorensen}\ \emph {et~al.}(1987)\citenamefont
  {Sorensen}, \citenamefont {Zhang},\ and\ \citenamefont
  {Taylor}}]{Sorensen1987}%
  \BibitemOpen
  \bibfield  {author} {\bibinfo {author} {\bibfnamefont {C.~M.}\ \bibnamefont
  {Sorensen}}, \bibinfo {author} {\bibfnamefont {H.~X.}\ \bibnamefont
  {Zhang}},\ and\ \bibinfo {author} {\bibfnamefont {T.~W.}\ \bibnamefont
  {Taylor}},\ }\bibfield  {title} {\bibinfo {title} {Cluster-size evolution in
  a coagulation-fragmentation system},\ }\href
  {https://doi.org/10.1103/PhysRevLett.59.363} {\bibfield  {journal} {\bibinfo
  {journal} {Physical Review Letters}\ }\textbf {\bibinfo {volume} {59}},\
  \bibinfo {pages} {363} (\bibinfo {year} {1987})}\BibitemShut {NoStop}%
\bibitem [{Note3()}]{Note3}%
  \BibitemOpen
  \bibinfo {note} {The homegeneity coefficients are defined as $K\left (\xi
  x,\xi y\right )=\xi ^\lambda K\left (x,y\right )$ and $F\left (\xi x,\xi
  y\right )=\xi ^\alpha F\left (x,y\right )$ for all $\left (x,y,\xi \right
  )\in \protect \mathbb {R}_+^3$.}\BibitemShut {Stop}%
\bibitem [{\citenamefont {Wattis}(2006)}]{Wattis2006}%
  \BibitemOpen
  \bibfield  {author} {\bibinfo {author} {\bibfnamefont {J.~A.~D.}\
  \bibnamefont {Wattis}},\ }\bibfield  {title} {\bibinfo {title} {{An
  introduction to mathematical models of coagulation–fragmentation processes:
  A discrete deterministic mean-field approach}},\ }\href
  {https://doi.org/10.1016/j.physd.2006.07.024} {\bibfield  {journal} {\bibinfo
   {journal} {Physica D}\ }\textbf {\bibinfo {volume} {222}},\ \bibinfo {pages}
  {1} (\bibinfo {year} {2006})}\BibitemShut {NoStop}%
\bibitem [{\citenamefont {Spicer}\ and\ \citenamefont
  {Pratsinis}(1996)}]{Spicer1996}%
  \BibitemOpen
  \bibfield  {author} {\bibinfo {author} {\bibfnamefont {P.~T.}\ \bibnamefont
  {Spicer}}\ and\ \bibinfo {author} {\bibfnamefont {S.~E.}\ \bibnamefont
  {Pratsinis}},\ }\bibfield  {title} {\bibinfo {title} {Coagulation and
  fragmentation: Universal steady-state particle-size distribution},\ }\href
  {https://doi.org/10.1002/aic.690420612} {\bibfield  {journal} {\bibinfo
  {journal} {AIChE Journal}\ }\textbf {\bibinfo {volume} {42}},\ \bibinfo
  {pages} {1612} (\bibinfo {year} {1996})}\BibitemShut {NoStop}%
\bibitem [{\citenamefont {Kryven}\ \emph {et~al.}(2014)\citenamefont {Kryven},
  \citenamefont {Lazzari},\ and\ \citenamefont {Storti}}]{Kryven2014}%
  \BibitemOpen
  \bibfield  {author} {\bibinfo {author} {\bibfnamefont {I.}~\bibnamefont
  {Kryven}}, \bibinfo {author} {\bibfnamefont {S.}~\bibnamefont {Lazzari}},\
  and\ \bibinfo {author} {\bibfnamefont {G.}~\bibnamefont {Storti}},\
  }\bibfield  {title} {\bibinfo {title} {{Population Balance Modeling of
  Aggregation and Coalescence in Colloidal Systems}},\ }\href
  {https://doi.org/10.1002/mats.201300140} {\bibfield  {journal} {\bibinfo
  {journal} {Macromolecular Theory and Simulations}\ }\textbf {\bibinfo
  {volume} {23}},\ \bibinfo {pages} {170} (\bibinfo {year} {2014})},\ \Eprint
  {https://arxiv.org/abs/https://onlinelibrary.wiley.com/doi/pdf/10.1002/mats.201300140}
  {https://onlinelibrary.wiley.com/doi/pdf/10.1002/mats.201300140} \BibitemShut
  {NoStop}%
\bibitem [{\citenamefont {Barthelmes}\ \emph {et~al.}(2003)\citenamefont
  {Barthelmes}, \citenamefont {Pratsinis},\ and\ \citenamefont
  {Buggisch}}]{Barthelmes2003}%
  \BibitemOpen
  \bibfield  {author} {\bibinfo {author} {\bibfnamefont {G.}~\bibnamefont
  {Barthelmes}}, \bibinfo {author} {\bibfnamefont {S.~E.}\ \bibnamefont
  {Pratsinis}},\ and\ \bibinfo {author} {\bibfnamefont {H.}~\bibnamefont
  {Buggisch}},\ }\bibfield  {title} {\bibinfo {title} {Particle size
  distributions and viscosity of suspensions undergoing shear-induced
  coagulation and fragmentation},\ }\href
  {https://doi.org/10.1016/S0009-2509(03)00133-7} {\bibfield  {journal}
  {\bibinfo  {journal} {Chemical Engineering Science}\ }\textbf {\bibinfo
  {volume} {58}},\ \bibinfo {pages} {2893} (\bibinfo {year}
  {2003})}\BibitemShut {NoStop}%
\bibitem [{\citenamefont {Delichatsios}\ and\ \citenamefont
  {Probstein}(1976)}]{Delichatsios1976}%
  \BibitemOpen
  \bibfield  {author} {\bibinfo {author} {\bibfnamefont {M.~A.}\ \bibnamefont
  {Delichatsios}}\ and\ \bibinfo {author} {\bibfnamefont {R.~F.}\ \bibnamefont
  {Probstein}},\ }\bibfield  {title} {\bibinfo {title} {{The Effect of
  Coalescence on the Average Drop Size in Liquid-Liquid Dispersions}},\ }\href
  {https://doi.org/10.1021/i160058a010} {\bibfield  {journal} {\bibinfo
  {journal} {Industrial \& Engineering Chemistry Fundamentals}\ }\textbf
  {\bibinfo {volume} {15}},\ \bibinfo {pages} {134} (\bibinfo {year} {1976})},\
  \Eprint {https://arxiv.org/abs/https://doi.org/10.1021/i160058a010}
  {https://doi.org/10.1021/i160058a010} \BibitemShut {NoStop}%
\bibitem [{\citenamefont {Kusters}(1991)}]{Kuster1991}%
  \BibitemOpen
  \bibfield  {author} {\bibinfo {author} {\bibfnamefont {K.~A.}\ \bibnamefont
  {Kusters}},\ }\emph {\bibinfo {title} {The influence of turbulence on
  aggregation of small particles in agitated vessels}},\ \href
  {https://doi.org/10.6100/IR362582} {\bibinfo {type} {phdthesis}},\ \bibinfo
  {school} {Chemical Engineering and Chemistry} (\bibinfo {year}
  {1991})\BibitemShut {NoStop}%
\bibitem [{\citenamefont {Sonntag}\ and\ \citenamefont
  {Russel}(1986)}]{Sonntag1986}%
  \BibitemOpen
  \bibfield  {author} {\bibinfo {author} {\bibfnamefont {R.~C.}\ \bibnamefont
  {Sonntag}}\ and\ \bibinfo {author} {\bibfnamefont {W.~B.}\ \bibnamefont
  {Russel}},\ }\bibfield  {title} {\bibinfo {title} {{Structure and Breakup of
  Flocs Subjected to Fluid Stresses: I. Shear Experiments}},\ }\href
  {https://doi.org/10.1016/0021-9797(87)90054-3} {\bibfield  {journal}
  {\bibinfo  {journal} {Journal of Colloid and Interface Science}\ }\textbf
  {\bibinfo {volume} {115}},\ \bibinfo {pages} {390} (\bibinfo {year}
  {1986})}\BibitemShut {NoStop}%
\bibitem [{\citenamefont {Sonntag}\ and\ \citenamefont
  {Russel}(1987{\natexlab{a}})}]{Sonntag1987}%
  \BibitemOpen
  \bibfield  {author} {\bibinfo {author} {\bibfnamefont {R.~C.}\ \bibnamefont
  {Sonntag}}\ and\ \bibinfo {author} {\bibfnamefont {W.~B.}\ \bibnamefont
  {Russel}},\ }\bibfield  {title} {\bibinfo {title} {{Structure and Breakup of
  Flocs Subjected to Fluid Stresses: II. Theory}},\ }\href
  {https://doi.org/10.1016/0021-9797(87)90053-1} {\bibfield  {journal}
  {\bibinfo  {journal} {Journal of Colloid and Interface Science}\ }\textbf
  {\bibinfo {volume} {115}},\ \bibinfo {pages} {378} (\bibinfo {year}
  {1987}{\natexlab{a}})}\BibitemShut {NoStop}%
\bibitem [{\citenamefont {Sonntag}\ and\ \citenamefont
  {Russel}(1987{\natexlab{b}})}]{Sonntag1987b}%
  \BibitemOpen
  \bibfield  {author} {\bibinfo {author} {\bibfnamefont {R.~C.}\ \bibnamefont
  {Sonntag}}\ and\ \bibinfo {author} {\bibfnamefont {W.~B.}\ \bibnamefont
  {Russel}},\ }\bibfield  {title} {\bibinfo {title} {{Structure and Breakup of
  Flocs Subjected to Fluid Stresses: III. Converging Flow}},\ }\href
  {https://doi.org/10.1016/0021-9797(86)90175-X} {\bibfield  {journal}
  {\bibinfo  {journal} {Journal of Colloid and Interface Science}\ }\textbf
  {\bibinfo {volume} {113}},\ \bibinfo {pages} {399} (\bibinfo {year}
  {1987}{\natexlab{b}})}\BibitemShut {NoStop}%
\bibitem [{\citenamefont {Harshe}\ \emph {et~al.}(2011)\citenamefont {Harshe},
  \citenamefont {Lattuada},\ and\ \citenamefont {Soos}}]{Harshe2011}%
  \BibitemOpen
  \bibfield  {author} {\bibinfo {author} {\bibfnamefont {Y.~M.}\ \bibnamefont
  {Harshe}}, \bibinfo {author} {\bibfnamefont {M.}~\bibnamefont {Lattuada}},\
  and\ \bibinfo {author} {\bibfnamefont {M.}~\bibnamefont {Soos}},\ }\bibfield
  {title} {\bibinfo {title} {{Experimental and Modeling Study of Breakage and
  Restructuring of Open and Dense Colloidal Aggregates}},\ }\href
  {https://doi.org/10.1021/la104658} {\bibfield  {journal} {\bibinfo  {journal}
  {Langmuir}\ }\textbf {\bibinfo {volume} {27}},\ \bibinfo {pages} {5739}
  (\bibinfo {year} {2011})}\BibitemShut {NoStop}%
\bibitem [{\citenamefont {Potanin}(1991)}]{Potanin1991}%
  \BibitemOpen
  \bibfield  {author} {\bibinfo {author} {\bibfnamefont {A.~A.}\ \bibnamefont
  {Potanin}},\ }\bibfield  {title} {\bibinfo {title} {{On the Mechanism of
  Aggregation in the Shear Flow of Suspensions}},\ }\href
  {https://doi.org/10.1016/0021-9797(91)90107-J} {\bibfield  {journal}
  {\bibinfo  {journal} {Journal of Colloid and Interface Science}\ }\textbf
  {\bibinfo {volume} {145}},\ \bibinfo {pages} {140} (\bibinfo {year}
  {1991})}\BibitemShut {NoStop}%
\bibitem [{\citenamefont {Potanin}(1992)}]{Potanin1992}%
  \BibitemOpen
  \bibfield  {author} {\bibinfo {author} {\bibfnamefont {A.~A.}\ \bibnamefont
  {Potanin}},\ }\bibfield  {title} {\bibinfo {title} {On the model of colloid
  aggregates and aggregating colloids},\ }\href
  {https://doi.org/10.1063/1.462229} {\bibfield  {journal} {\bibinfo  {journal}
  {Journal of Chemical Physics}\ }\textbf {\bibinfo {volume} {96}},\ \bibinfo
  {pages} {9191} (\bibinfo {year} {1992})}\BibitemShut {NoStop}%
\bibitem [{\citenamefont {Potanin}\ and\ \citenamefont
  {Russel}(1996)}]{Potanin1996}%
  \BibitemOpen
  \bibfield  {author} {\bibinfo {author} {\bibfnamefont {A.~A.}\ \bibnamefont
  {Potanin}}\ and\ \bibinfo {author} {\bibfnamefont {W.~B.}\ \bibnamefont
  {Russel}},\ }\bibfield  {title} {\bibinfo {title} {{Fractal model of
  consolidation of weakly aggregated colloidal dispersions}},\ }\href
  {https://doi.org/10.1103/PhysRevE.53.3702} {\bibfield  {journal} {\bibinfo
  {journal} {Physical Review E}\ }\textbf {\bibinfo {volume} {53}},\ \bibinfo
  {pages} {3702} (\bibinfo {year} {1996})}\BibitemShut {NoStop}%
\bibitem [{\citenamefont {Gibaud}\ \emph
  {et~al.}(2020{\natexlab{a}})\citenamefont {Gibaud}, \citenamefont {Dag\`es},
  \citenamefont {Lidon}, \citenamefont {Jung}, \citenamefont {Ahour\'e},
  \citenamefont {Sztucki}, \citenamefont {Poulesquen}, \citenamefont {Hengl},
  \citenamefont {Pignon},\ and\ \citenamefont {Manneville}}]{Gibaud2020a}%
  \BibitemOpen
  \bibfield  {author} {\bibinfo {author} {\bibfnamefont {T.}~\bibnamefont
  {Gibaud}}, \bibinfo {author} {\bibfnamefont {N.}~\bibnamefont {Dag\`es}},
  \bibinfo {author} {\bibfnamefont {P.}~\bibnamefont {Lidon}}, \bibinfo
  {author} {\bibfnamefont {G.}~\bibnamefont {Jung}}, \bibinfo {author}
  {\bibfnamefont {L.~C.}\ \bibnamefont {Ahour\'e}}, \bibinfo {author}
  {\bibfnamefont {M.}~\bibnamefont {Sztucki}}, \bibinfo {author} {\bibfnamefont
  {A.}~\bibnamefont {Poulesquen}}, \bibinfo {author} {\bibfnamefont
  {N.}~\bibnamefont {Hengl}}, \bibinfo {author} {\bibfnamefont
  {F.}~\bibnamefont {Pignon}},\ and\ \bibinfo {author} {\bibfnamefont
  {S.}~\bibnamefont {Manneville}},\ }\bibfield  {title} {\bibinfo {title}
  {{Rheoacoustic Gels: Tuning Mechanical and Flow Properties of Colloidal Gels
  with Ultrasonic Vibrations}},\ }\href
  {https://doi.org/10.1103/PhysRevX.10.011028} {\bibfield  {journal} {\bibinfo
  {journal} {Physical Review X}\ }\textbf {\bibinfo {volume} {10}},\ \bibinfo
  {pages} {1} (\bibinfo {year} {2020}{\natexlab{a}})},\ \bibinfo {note}
  {011028},\ \Eprint {https://arxiv.org/abs/1905.07282} {1905.07282}
  \BibitemShut {NoStop}%
\bibitem [{\citenamefont {Gibaud}\ \emph
  {et~al.}(2020{\natexlab{b}})\citenamefont {Gibaud}, \citenamefont {Divoux},\
  and\ \citenamefont {Manneville}}]{Gibaud2020b}%
  \BibitemOpen
  \bibfield  {author} {\bibinfo {author} {\bibfnamefont {T.}~\bibnamefont
  {Gibaud}}, \bibinfo {author} {\bibfnamefont {T.}~\bibnamefont {Divoux}},\
  and\ \bibinfo {author} {\bibfnamefont {S.}~\bibnamefont {Manneville}},\
  }\bibinfo {title} {{Encyclopedia of Complexity and Systems Science}}\
  (\bibinfo  {publisher} {Springer},\ \bibinfo {year} {2020})\ Chap.\ \bibinfo
  {chapter} {Nonlinear mechanics of colloidal gels: creep, fatigue and
  shear-induced yielding}, pp.\ \bibinfo {pages} {1--24}\BibitemShut {NoStop}%
\bibitem [{\citenamefont {Dag\'es}\ \emph {et~al.}(2021)\citenamefont
  {Dag\'es}, \citenamefont {Lidon}, \citenamefont {Jung}, \citenamefont
  {Pignon}, \citenamefont {Manneville},\ and\ \citenamefont
  {Gibaud}}]{Dages2021}%
  \BibitemOpen
  \bibfield  {author} {\bibinfo {author} {\bibfnamefont {N.}~\bibnamefont
  {Dag\'es}}, \bibinfo {author} {\bibfnamefont {P.}~\bibnamefont {Lidon}},
  \bibinfo {author} {\bibfnamefont {G.}~\bibnamefont {Jung}}, \bibinfo {author}
  {\bibfnamefont {F.}~\bibnamefont {Pignon}}, \bibinfo {author} {\bibfnamefont
  {S.}~\bibnamefont {Manneville}},\ and\ \bibinfo {author} {\bibfnamefont
  {T.}~\bibnamefont {Gibaud}},\ }\bibfield  {title} {\bibinfo {title}
  {Mechanics and structure of carbon black gels under high-power ultrasound},\
  }\href {https://doi.org/10.1122/8.0000187} {\bibfield  {journal} {\bibinfo
  {journal} {Journal of Rheology}\ }\textbf {\bibinfo {volume} {65}},\ \bibinfo
  {pages} {477} (\bibinfo {year} {2021})}\BibitemShut {NoStop}%
\bibitem [{\citenamefont {Klimchitskaya}\ \emph {et~al.}(2000)\citenamefont
  {Klimchitskaya}, \citenamefont {Mohideen},\ and\ \citenamefont
  {Mostepanenko}}]{Klimchitskaya2000}%
  \BibitemOpen
  \bibfield  {author} {\bibinfo {author} {\bibfnamefont {G.~L.}\ \bibnamefont
  {Klimchitskaya}}, \bibinfo {author} {\bibfnamefont {U.}~\bibnamefont
  {Mohideen}},\ and\ \bibinfo {author} {\bibfnamefont {V.~M.}\ \bibnamefont
  {Mostepanenko}},\ }\bibfield  {title} {\bibinfo {title} {{Casimir and Van der
  Waals force between two plates or a sphere (lens) above a plate made of real
  metals}},\ }\href {https://doi.org/10.1103/PhysRevA.61.062107} {\bibfield
  {journal} {\bibinfo  {journal} {Physical Review A}\ }\textbf {\bibinfo
  {volume} {61}},\ \bibinfo {pages} {062107} (\bibinfo {year}
  {2000})}\BibitemShut {NoStop}%
\bibitem [{\citenamefont {Visser}(1972)}]{Visser1972}%
  \BibitemOpen
  \bibfield  {author} {\bibinfo {author} {\bibfnamefont {J.}~\bibnamefont
  {Visser}},\ }\bibfield  {title} {\bibinfo {title} {{On Hamaker constants: A
  comparison between Hamaker constants and Lifshitz-Van der Waals constants}},\
  }\href {https://doi.org/10.1016/0001-8686(72)85001-2} {\bibfield  {journal}
  {\bibinfo  {journal} {Advances in Colloid and Inferface Science}\ }\textbf
  {\bibinfo {volume} {3}},\ \bibinfo {pages} {331} (\bibinfo {year}
  {1972})}\BibitemShut {NoStop}%
\bibitem [{\citenamefont {Waite}\ \emph {et~al.}(2001)\citenamefont {Waite},
  \citenamefont {Cleaver},\ and\ \citenamefont {Beattie}}]{Waite2001}%
  \BibitemOpen
  \bibfield  {author} {\bibinfo {author} {\bibfnamefont {T.~D.}\ \bibnamefont
  {Waite}}, \bibinfo {author} {\bibfnamefont {J.~K.}\ \bibnamefont {Cleaver}},\
  and\ \bibinfo {author} {\bibfnamefont {J.~K.}\ \bibnamefont {Beattie}},\
  }\bibfield  {title} {\bibinfo {title} {{Aggregation Kinetics and Fractal
  Structure of $\gamma$-Alumina Assemblages}},\ }\href
  {https://doi.org/10.1006/jcis.2001.769} {\bibfield  {journal} {\bibinfo
  {journal} {Journal of Colloid and Interface Science}\ }\textbf {\bibinfo
  {volume} {241}},\ \bibinfo {pages} {333} (\bibinfo {year}
  {2001})}\BibitemShut {NoStop}%
\end{thebibliography}%

\end{document}